\documentclass[11pt, a4paper]{article}

\usepackage{amssymb}
\usepackage{amsmath}
\usepackage{amsfonts}
\usepackage{array}
\usepackage{epsfig}
\usepackage{fancyhdr}
\usepackage{booktabs}
\usepackage{lastpage}
\usepackage{siunitx}

\usepackage[usenames,dvipsnames]{color}
\usepackage[hyperindex, linktocpage, hidelinks]{hyperref}
\usepackage{inputenc} 

\usepackage{fullpage}
\usepackage{multirow}
\usepackage{floatflt}
\usepackage{graphicx}
\usepackage{caption}
\usepackage{subcaption}
\usepackage{gensymb}
\usepackage{multirow}
\usepackage{makecell}
\usepackage{longtable}
\usepackage{tabu}
\usepackage[svgnames]{xcolor}
\usepackage{float} 
\usepackage{listings} 
\usepackage{bookmark} 
\usepackage{braket} 
\usepackage{wrapfig}
\usepackage{footnote}
\usepackage{comment}
\usepackage{setspace}
\usepackage{graphicx}
\usepackage{ifplatform}
\usepackage{pst-pdf}
\usepackage{xkeyval}
\usepackage{mathtools}

\usepackage{chemformula}
\usepackage{tikz}
\usepackage{mathptmx}
\usepackage{geometry}
\usepackage{pstricks}
\usepackage{indentfirst} 
\usepackage[T1]{fontenc}

\usepackage{titlesec}
\usepackage{authblk}

\usepackage{colortbl}
\definecolor{LightGreen}{rgb}{0.718, 0.98, 0.604}
\definecolor{DarkGreen}{rgb}{0.149, 0.439, 0.024}

\newcommand{\bc}{\begin{cases}\begin{aligned}}
\newcommand{\ec}{\end{aligned}\end{cases}}
\newcommand{\eq}{\begin{equation}}
\newcommand{\fine}{\end{equation}}
\newcommand{\beq}{\begin{equation}}
\newcommand{\eeq}{\end{equation}}

\newcommand{\uno}{\leavevmode\hbox{\small1\normalsize\kern-.33em1}}

\newcommand{\casi}{\begin{cases}\begin{aligned}}
\newcommand{\casiend}{\end{aligned}\end{cases}}

\newcommand{\x}{{\bf x}}

\newcommand{\GB}{{\rm GB}(\x,z)}

\titleformat*{\section}{\large\upshape}
\titleformat*{\subsection}{\large\upshape}
\titleformat*{\subsubsection}{\normalsize\upshape} 

\usetikzlibrary{
  arrows,
  shapes,
  snakes,
  fit,
  shapes.misc,
  shapes.arrows,
  shapes.geometric,
  chains,
  matrix,
  positioning,
  scopes,
  decorations.pathmorphing,
  shadows,
  calc,
  fadings,
  decorations.pathreplacing
}

\renewcommand{\arraystretch}{1.3} 

\title{\centering 
  The {\sl Tree of Light} as interstellar optical transmitter system}

      \author[1]{Elisa Bazzani}
      \author[1]{Anna Valeria Guglielmi}
      \author[1]{Roberto Corvaja}
      \author[1]{Nicola Laurenti}
      \author[2,3]{Filippo Romanato}
      \author[2,3]{Gianluca Ruffato}
      \author[2,3]{Andrea Vogliardi}
      \author[1,3]{Francesco Vedovato}
      \author[1,2,3]{Giuseppe Vallone}
      \author[1]{Lorenzo Vangelista}
      \author[1,3]{Paolo Villoresi}%

  {
    \affil[1]{Dept. Information Engineering - University of Padova, v. Gradenigo 6B, 35131 Padua, Italy,} 
    \affil[2]{Dept. Physics and Astronomy - University of Padova, v. Marzolo 8, 35131 Padua, Italy,}
    \affil[3]{Padua Quantum Technologies Research Center, University of Padova}
  }%
\date{\ }

\begin{document}
\maketitle
The hunt for habitable planets outside the solar system as well as the search for the evidence of extraterrestrial life are everlasting questions for humanity to ponder. About the first aspect, concrete scientific evidences have grown steadily in the past decades. The discoveries of extrasolar planets with habitable conditions similar to the Earth started in 2007, with the observation of Gliese 581c using the observations made from La Silla (Chile) with the HARPS spectrograph on the ESO 3.6-m telescope \cite{Udry7}. The observations started with ground telescopes and expanded including space telescopes with several space missions starting from Kepler and including CHEOPS, GAIA, TESS and others as well as the coming PLATO mission.

In addition to the observation from the Earth or its close surrounding, the trip to the vicinity of an extrasolar planet for direct observations has been conceived. In particular, the Starshot Project supported by the Breakthrough Initiatives is developing the conceptual study on the feasibility of a trip aiming at the $\alpha$-Centauri star system, the closest candidate \cite{BIStarshot}. More than one  exoplanets are orbiting within the habitable zone of star Proxima Centauri, a red dwarf star member of the three-star system, including the exoplanet Proxima Centauri b. In order to cover the 4.2 light-years of separation in trip lasting 20 years, a very lightweight probe is considered, suitable to be accelerated to 20\% of the light-speed by directed  energy propulsion, in line with the visionary proposal by R. L. Forward in 1984 \cite{Forw}. 
The probe is conceived then as a sail that is pushed by a so-called photon engine, based on the coherent combination of laser beams. All the systems needed for the navigation, observation and communications of the findings shall be located on the sail surface, including the detectors for the acquisition of the local information of the mission,  the processing unit, the memory, the signal generator and the optical transmitter to send the collected information to Earth. The optical signal is intended to be received by an array of telescopes, with single-photon sensitivity. The general communication system and the assessment of the data rate have been already the subject of detailed studies \cite{Messerschmitt_2020}.

In this context, this work aims at investigating the optical transmission system needed for such lightweight sail, taking  into account the physical constraints of such unprecedented link and focusing on  the  optimal scheme for the optical signal emission. In particular,  the optical signal is distributed  to several emitters on the sail. The light diffraction resulting from the pattern of the emitters acting coherently determines the characteristics of the whole beam transmitted by the sail and of the received signal on the Earth. The performance of the digital communication system using pulse position modulation (PPM) can be assessed and channel coding schemes are proposed. 

\begin{figure}[H]
\begin{center}
\includegraphics[width=11 cm]{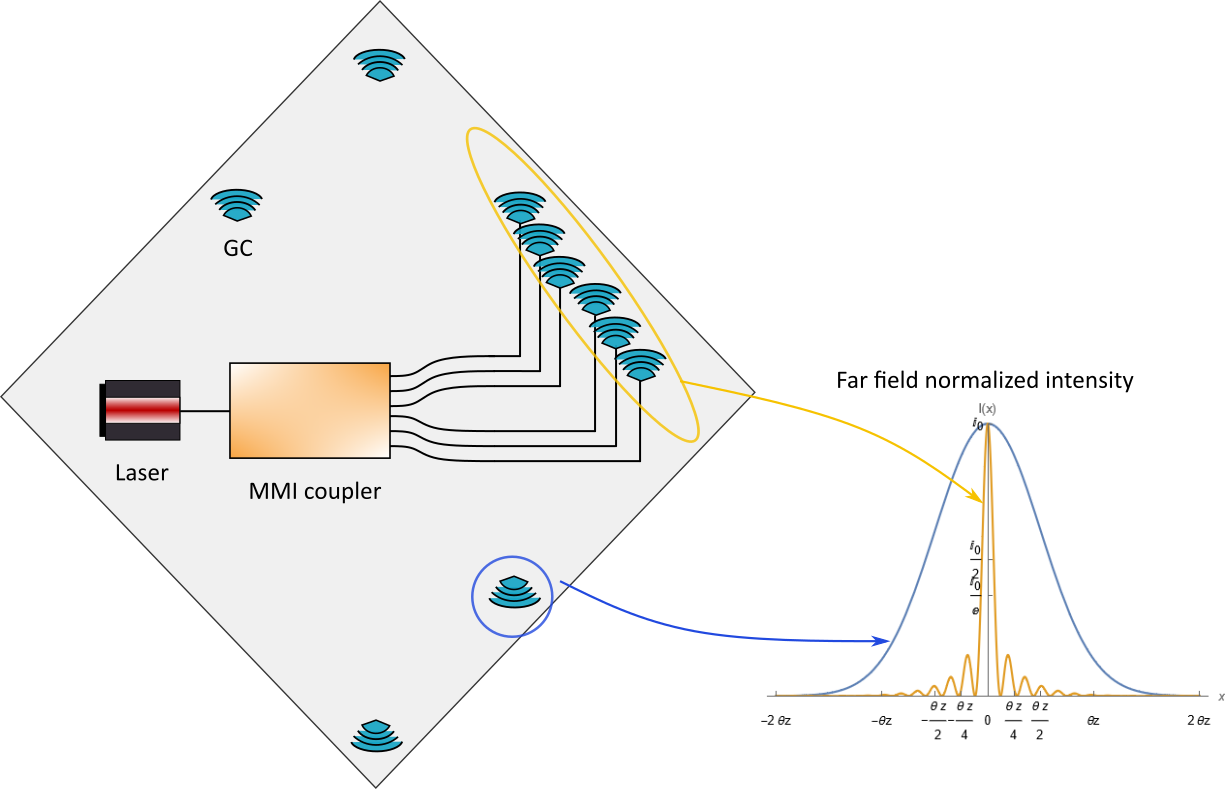}
\caption{Tree-of-Light working principle: the seed laser power, coming from the tree trunk, is divided into branches by means of a Multi-mode interference (MMI) coupler, which delivers the signals to $N$ leafs,  that are grating couplers, not to scale. The individual GC couples the fundamental Gaussian mode (blue), while the array, by exploiting the GCs interferences, gives rise to a narrower main lobe (yellow).}\label{fig:qcToL}
\end{center}
\end{figure}

We are using the paradigm for which the entire sail communication system is described as a Tree-of-light (ToL), sketched in Fig.~\ref{fig:qcToL}: the detectors, CPU, memory and laser transmitter are the central unit, representing the trunk of the tree. The  branches of the tree are the waveguides, directed to the sail surface. By means of multimode splitters,  the signal is further distributed via the petioles to the emitters, the leaves, realized by grating couplers (GCs), on which this work is more focused. 

In Section~\ref{sec:concept} the concept of the ToL will be detailed, first with the analysis of the the conditions in which the Starshot sail transmitter will operate, considering the constraints in mass and the hypotheses on the available power. 
Then, the emitters scheme will be addressed, with a focus on the effects of the pattern of emitters to realize an Optical Phased Array (OPA), in order to restrict and possibly steer the main emission lobe to the desired pointing angle. 
The expected link losses are evaluated with different ToL configurations, in order to assess the photon rate at the receiver. 

In the following Section~\ref{sec:sailident} we present a proposal for the identification of the sails based on the actual Doppler shift experienced by the sails in the acceleration phase, which can be considered random around a mean value. The probability of distinguishing the sails by this technique is evaluated.

In Section~\ref{sec:lbdr}, the digital communications aspects using PPM are addressed, on the base of the previous results on the expected photon flux at the receiver.
In particular the channel coding to implement the error correction is studied. Among the possible channel coding strategies for the Poisson channel,  serially-concatenated PPM (SCPPM) schemes and Low-Density Parity Check (LDPC) codes are considered and their performance is presented.

Finally, in Section~\ref{sec:leaves}, the design and realization of the grating couplers on the surface of the sail is discussed. 
GCs are motivated for keeping the size and weight of the transmitter as small as possible, since they are realized directly on the surface of the sail, similarly to the waveguides of  branches and petioles. Morever, this technique can benefit from the parallelization that has already been  demonstrated at the technological level for the production of nano-optics on an industrial scale, reducing the cost per single sail.
The size and the phase front quality of the mode  emitted by the GCs are crucial parameters as they define the divergence of the beam emitted by the sail toward the receiver, having an impact on the photon flux as discussed in Section~\ref{sec:concept}.
\section{Concept  of the ToL optical transmitter}\label{sec:concept}


The optical scheme is  crucial  for the sail communication systems, since the diffraction losses derive from it. These are the dominant term in the photon budget, due to the small extension of  the available receiving area of the Starshot receiving telescope,  or array of telescopes  with respect to  the large size of the beam that carry the data payload, which are the  information acquired by the sail and encoded by means of optical pulses. 


As schematized in Fig.~\ref{fig:qcToL}, the individual GC emits a relatively large beam, shown with the blue curve, in the range of few to few tens of mrad, while the effect of the array is to shrink by more than two orders of magnitude the main lobe following the calculations reported below, and giving the yellow curve.  



\subsection{Context and main optical losses for the sail transmission from $\alpha$-Centauri}

The Starshot sail optical transmission shall operate once  reached    $\alpha$-Centauri star system and after the acquisition of  the relevant information to communicate by the onboard sensors.

\begin{figure}[H]
\begin{center}
\includegraphics[width=15 cm]{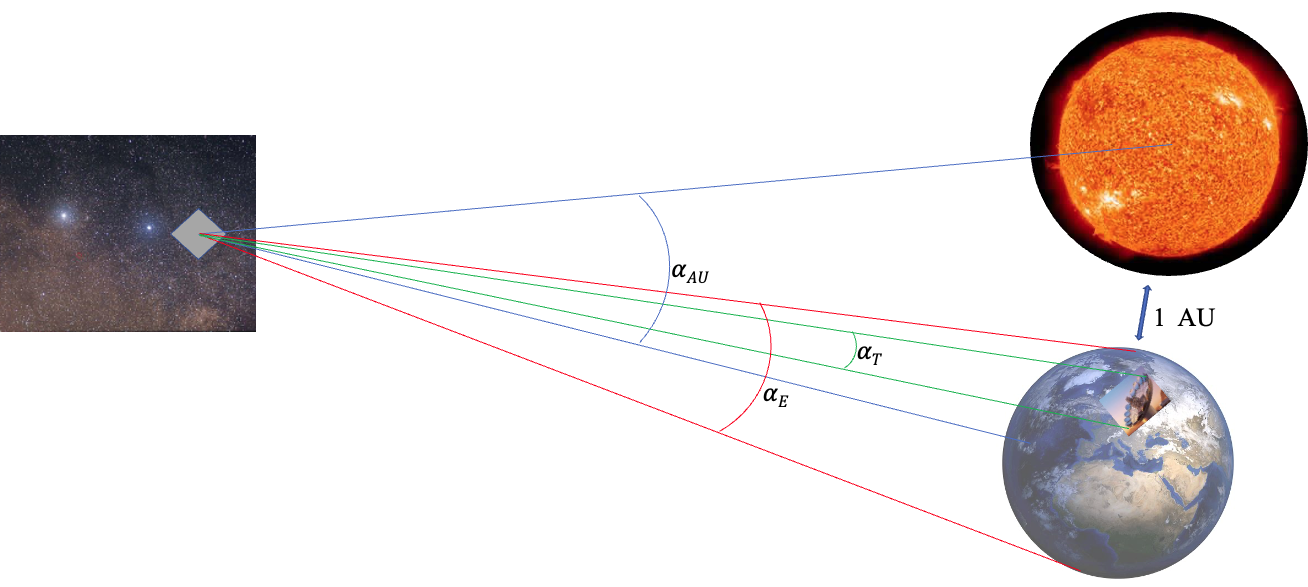}
\caption{Definition of the relevant angles in the ToL communication scheme - not to scale.}\label{fig:ToLangle}
\end{center}
\end{figure}

With reference to Fig. \ref{fig:ToLangle}, the  $\alpha$-Centauri star system is separated from the Solar system by about 4.4 ly, that are 4.1 $10^{16}$m. 

The angle at which one astronomical unis (AU = 1.496 $10^{11}$m) is subtended from there is $\alpha_{AU}$ = 3.62 $\mu$rad. 

The angle at which the Earth diameter (Earth radius = 6.378 $10^{6}$m) is subtended from there is $\alpha_{E}$ = 0.31 nrad. 

The subtended angle at which a ground receiver based on an array of telescopes whose size is supposed to be of one kilometer is $\alpha_{T}$ = 24.2 femtorad (24.2 $10^{-15}$rad).

By considering the 1 km size telescope array as the candidate Starshot receiver, the loss due to the overlap of a beam as large as the angle spanning the Earth may be assessed as  $Loss_E =  ( \frac{\alpha_{T}}{\alpha_E})^2 = 6.1$ $10^{-9}$ while the loss in the case of a beam spanning 1 AU results of $Loss_{AU} =  (\frac{\alpha_{T}}{\alpha_{AU}})^2 = 4.5$ $ 10^{-17}$.

The photon budget is also affected by other factors as the atmospheric turbulence and attenuation, background light and detector noise, not analysed in this work, but it is mainly constrained by these harsh losses imposed by the extreme distance of the sail, the limitation in the size and weight of the transmitter and the restriction to the optical spectral region. 

From such quick assessments, we may derive the correspondence between the  divergence that the sail optical transmitter is imposing to the optical beam and  the level of losses resulting from the coupling to the receiver telescopes. We here consider the microradian divergence value as a reasonable compromise. Indeed, the size of the sail and the corresponding allowed array size as well as the limitation in the GCs phasing to achieve a optimal coherent emission from a ultra lightweight structure as the Starshot sail are pointing at a reasonable coherent extent of the order of a meter or a large fraction of it, corresponding to about a span of a million wavelengths.  With it, the diffraction losses results to be  over 16 orders of magnitude. We note that this estimate is larger than the overall attenuation of the first experimental single photon exchange for quantum communications with a satellite, reported by Villoresi et al. in 2008, which was determined as $-157$\,dB \cite{Villoresi2008}. 
However, even if in that experiment the source  average power was of 8.3 mW, the average count rate was observed to be  of 5 clicks-per-second (cps). This results follows  from the relatively low energy of each photon: the green photons at $\lambda=$ 532 nm used there have an energy  $E= h \frac{c}{\lambda} = $ 3.7 $10^{-19}$ J. A similar argument may be appled in the case of the Starshot photon budget, in which, considering also a reasonable values for the other causes for losses mentioned above, a minimum average optical power of the order of 1 W is needed to be emitted by the sail.

We conclude this initial analysis mentioning that the  study carried out by Berera and Calderón-Figueroa and envisaging the use of the transmission in X-ray band would provide much lower losses by the communication system \cite{BCFig}. However we consider here that the  feasibility of the entire system in this band is  beyond reach.

\subsection{Optical phased array of transmitting leaves}
OPAs are used to combine the emission of many GCs to increase the on-axis intensity by constructive interference as well as for precise steering of the beam lobe \cite{Guo2021}. Indeed, the many light-year distance separating the transmitter from the Earth receiver requires an unprecedented beam forming in a narrow lobe and a corresponding pointing precision. 
However, the use of beam steering by the rigid rotating the sail is not a functionality considered as feasibe so far.

\begin{figure}[H]
\begin{center}
\includegraphics[width=15 cm]{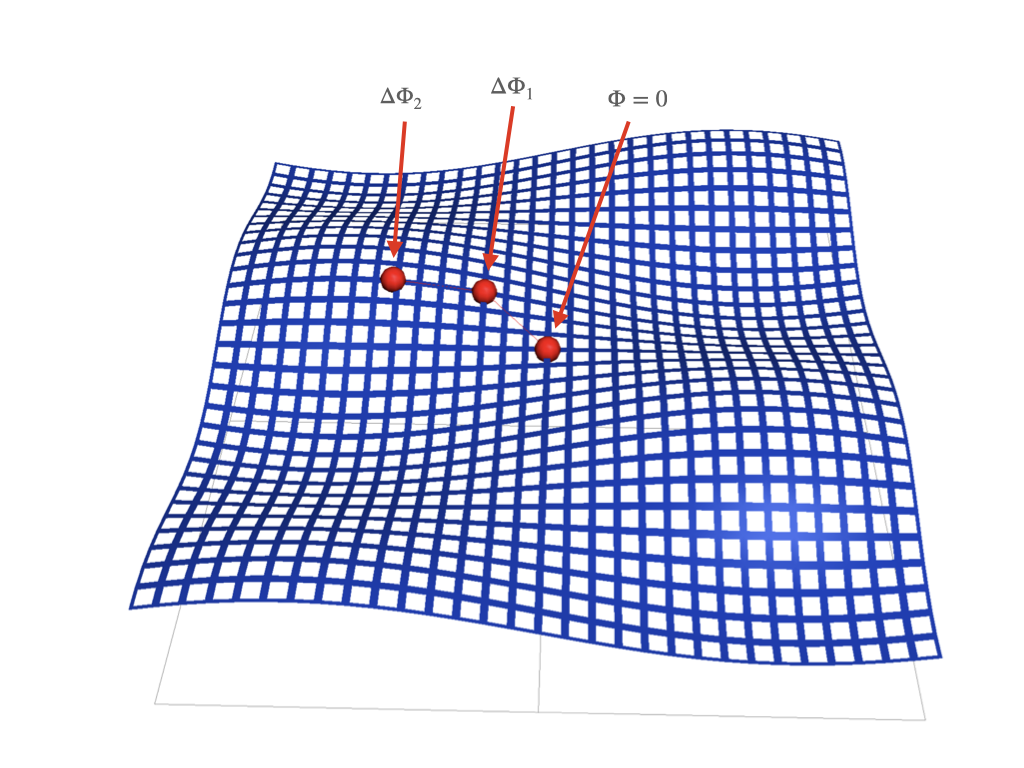}
\caption{Sail surface deformation and  phase variation in two points with respect to a reference due to surface deformation. To achieve the coherent emission for all GCs at the grid nodes, an interpolation will be used.}\label{fig:sailOPA}
\end{center}
\end{figure}

To achieve the coherent combination of the GCs emission, the knowledge of the  phase difference of the different points on the  sail surface at the emission time,  as shown in Fig. \ref{fig:sailOPA}, and its correction are needed to be implemented. 

Indeed, the sail actual geometry results from the forces exerted at the acceleration and from the forces due to the particle absorption during the interstellar flight and the subsequent  dynamical evolution. The determination of the phase to be added to any GC with respect to a reference one may be determined by means of a feedback from the beam splitters used to distribute the optical signal to the GCs, as described below.

Once the phase correction is determined, a phase modulation stage after the beam subdivision will be used for the signal directed to the GCs.

OPA elements can be  fabricated  separately and then integrated to form the entire array\cite{Wolfgang1994}.  Similar approaches explointing integrated waveguide optics have been already proposed and implemented by using AlGaAs\cite{Vasey1993} and silicon\cite{VanAcoleyen2009} and other \cite{Zhang21} technologies. \cite{VanAcoleyen2009} in particular characterize a silicon-on-insulator integrated circuit, that includes a MMI splitting system, a modulation stage and a 1D grating couplers output array, an implementation thus comparable to the ToL concept here described.

These integrated optical technologies provides concrete means for the harnessing of the sail optical system, including the scaling capacity. OPAs have been therefore identified as a versatile and lightweight solution to adjust small misalignments of the transmitter, so as to increase the photon budget at the receiver side.


\subsection{Diffraction model of Gaussian aperture systems}


We briefly recall the model based on diffraction theory for an ensemble of identical coherent emitters. The individual element, i.e. GC, 
is considered to couple in free-space a fundamental Gaussian mode.

The complex function $GB( {\bf x},t)$ describing a Gaussian beam propagating along the $z$ axis and centered in $\x_0=0$ in the transverse plane and at the distance $d_0$ from the origin, is expressed as

\eq
\label{Gbeam}
U_\omega(\vec x,t)=\GB e^{i\omega (t-\frac1cz)}\,,
\qquad
\GB=\sqrt{\frac{kz_0 P_0}{\pi}}\frac{i}{q_z}e^{-ik\frac{(\x-\x_0)^2}{2q_z}}
\fine
with
\eq
k=\frac{\omega}{c}\,,\qquad\qquad q_z=z+q_0=z-d_0+iz_0
\fine
In the far field ($z\gg w_0$), the field is equivalent to its Fourier transform evaluated in ${\bf k}=k\frac{\x}{z}$. Therefore
\eq
\GB_{\x_0}\simeq\frac{w_0 k}{z}\sqrt{\frac{P_0}{2\pi}}e^{i\frac{k q_z}{2z^2}\x^2}e^{i\frac{k\x\cdot\x_0}{z}}
=\frac{w_0 k}{z}\sqrt{\frac{P_0}{2\pi}}e^{i\frac{k\x^2}{2z}}e^{-\frac{k z_0}{2z^2}\x^2}e^{i\frac{k\x\cdot\x_0}{z}}
\fine
Let us assume to distribute equally the total emitted optical power $ P_0$ among $N$ equivalent GCs, located  on different positions $\x_{0j}$ of the same plane, orthogonal with respect to the direction to the receiver, with $j=0,\cdots N-1$. The field produced by each sub aperture has then power $P_0/N$, negletting the splitting losses. 

$P_0$ is assumed to be the optical power available for transmission, regardless to the design of the light source, which can be a single laser as well as multiple phase-locked or injection-locked lasers\cite{Heck2017}, and after the attenuation introduces in the optical harnessing along the ToL trunk, branches and leaves. 

The far-field amplitude is then given by

\eq
G(\x,z)\simeq
N\frac{w_0 k}{z}\sqrt{\frac{P_0}{2\pi N}}e^{i\frac{k\x^2}{2z}}e^{-\frac{k z_0}{2z^2}\x^2}F_N(\x) 
\label{eq:G}
\fine
where we have multiplied and divided by $N$ for convenience and

\beq
F_N(\x)=\frac1N\sum_{j=0}^{N-1}e^{\frac{i k}{z}\x\cdot \x_{0j}}	\,,\qquad F_N(\x=0)=1
\label{eq:AF}
\eeq
is the important factor arising from the interference among the emitters.

Since $z_0=\frac{kw_0^2}{2}$, the intensity in far-field is given by
\eq
I(\x,z)=|G(\x,z)|^2=
I_0(z) e^{-(\frac{\pi w_0}{\lambda z})^2\x^2}|F_N(\x)|^2
\label{eq:ffpatterntot}
\fine
namely a Gaussian envelope modulated by the $|F_N(\x)|^2$ function. 


In the previous equation we defined the on-axis intensity by
\eq
I_0(z)=I(0,z)=
\frac{Nw^2_0 k^2P_0}{2\pi z^2}
\label{eq:qc_central_int}
\fine
$I_0$ is proportional to $N w_0^2$, namely the total ``effective area" $A_{eff}^{Tx}$ of the transmitter, which can be increased by increasing $N$ or the size of each single beam. Eq. \ref{eq:qc_central_int} also expresses the linear relation between the total instantaneous power $P_0$ at the source and the on-axis intensity in the far field.

\begin{figure}[H]
\begin{center}
\includegraphics[width=13 cm]{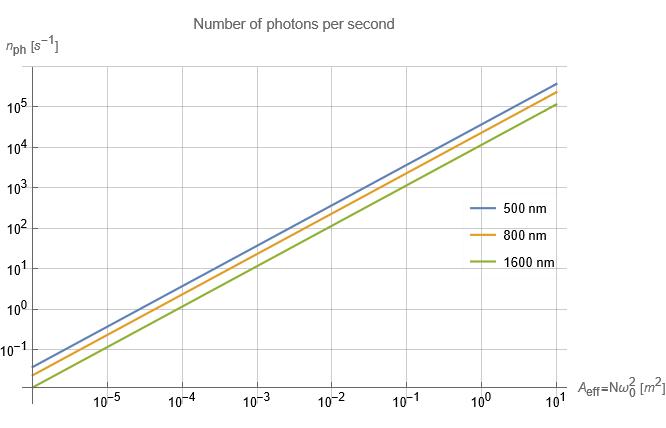}
\caption{Photon rate at the receiver as a function of the transmitter effective area for different wavelengths.}\label{fig:qcNumPhAeff}
\end{center}
\end{figure}

Without including other attenuation and finite efficiencies, as stated  above already, we may estimate the maximum number of signal photons per second: the photon rate, collected by a receiver of effective area $A_{eff}^{Rx}$, is then given by

\eq
n_{ph}(z) = \frac{I_0(z) A_{eff}^{Rx}}{E_{ph}} = \frac{I_0(z) A_{eff}^{Rx}\lambda}{h c}
\label{eq:qc_ph_rate}
\fine
where $E_{ph} = hc/\lambda$ is the photon energy.
By assuming a perfect pointing, that allows to use the maximum intensity value, that is found on the beam axis, Fig. \ref{fig:qcNumPhAeff} shows $n_{ph}$ as a function of the transmitter effective area, for three different wavelengths.
To evaluate $n_{ph}$, we assumed a square receiver with $A_{eff}^{Rx}$ = 1 km$^2$, $P_0=1$ W and $\lambda = 800$ nm.

\subsubsection{Square lattice Optical-Phased-Array}
\begin{figure}[H]
\centering
\begin{subfigure}[H]{.25\textwidth} 
\centering
  \includegraphics[width= 4 truecm]{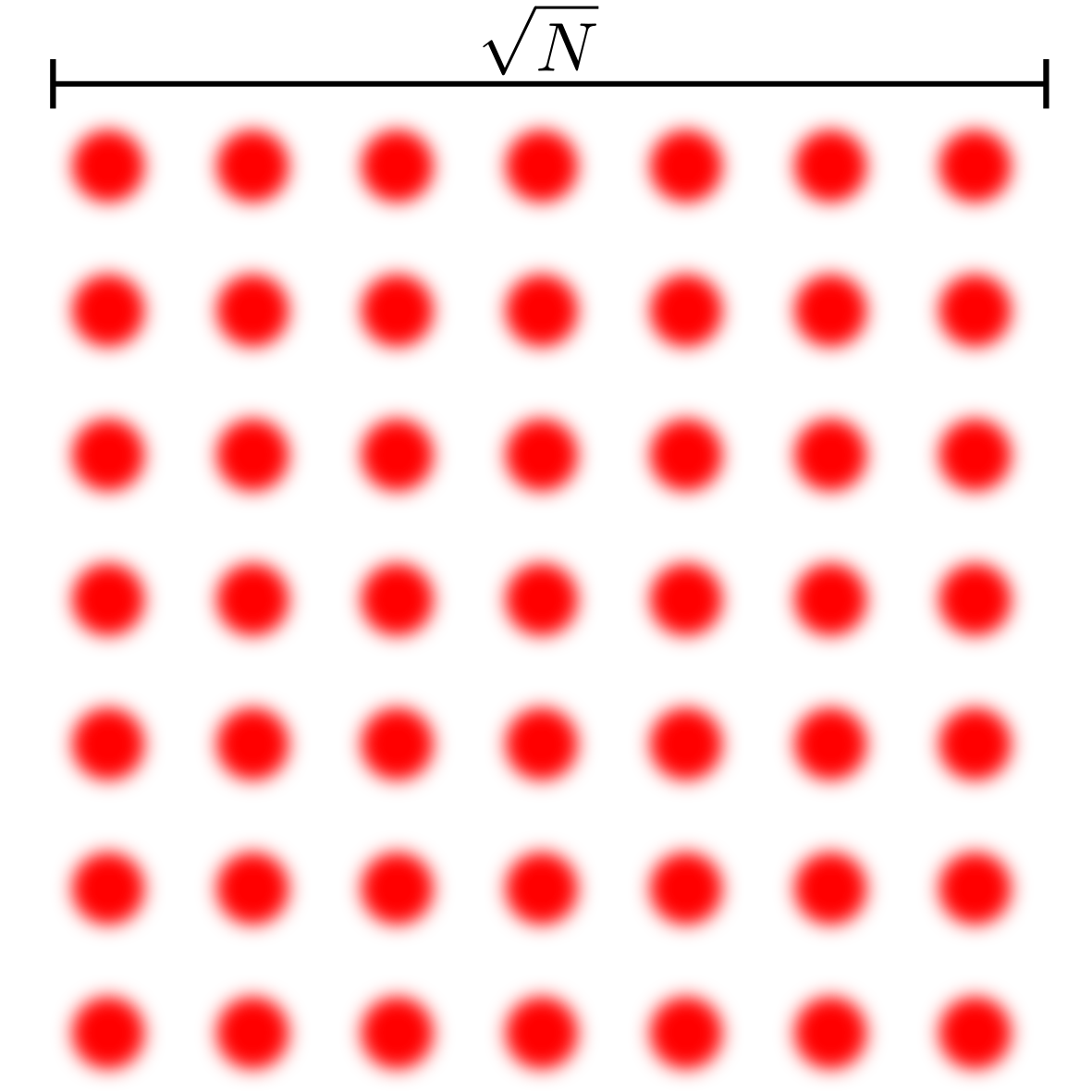}
  \caption{Beam center locations of the lattice array.}
  \label{qc:lattice}
\end{subfigure} 
\hskip .1 truecm
\begin{subfigure}[H]{.7\textwidth}
\centering
  \includegraphics[width= 9 truecm]{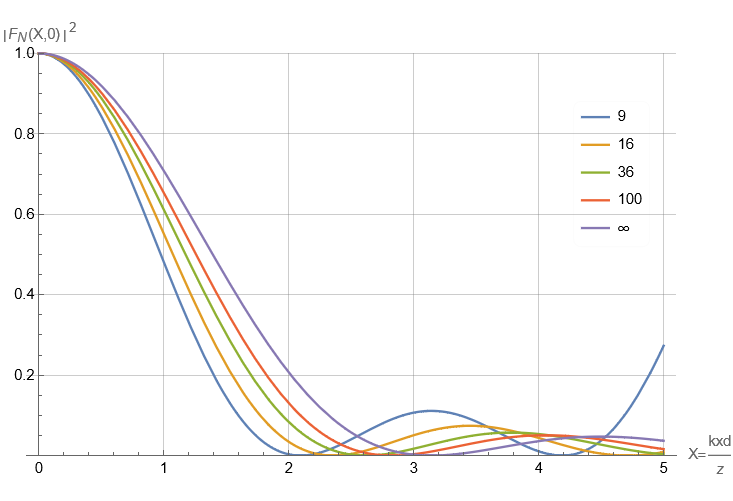}
  \caption{$|F_N|^2$ of the lattice array for different $N$ beams centers.}
  \label{qc:FNl}
\end{subfigure}
\label{}
\caption{Lattice array arrangement and squared array factor.}
\end{figure}

According to Eq. (\ref{eq:AF}), the interference term depends on the emitter spatial distribution. Here we describe the array factor for a lattice array of  $N$ emitters, taken as a square number, with a  side  of length $2d$ that is formed by $\sqrt{N}$ equally distributed beam centers as depicted in Fig. \ref{qc:lattice}.
If $(0,0)$ is the coordinate of the lattice center, the locations of the beam center in the h-th row and j-th column is given by
\beq
\begin{aligned}
x_{hj}&=-d+\frac{2d}{\sqrt N-1}h\,,\qquad h=0,\cdots \sqrt N-1
\\
y_{hj}&=-d+\frac{2d}{\sqrt N-1}j\,,\qquad j=0,\cdots \sqrt N-1
\end{aligned}
\label{eq:coordlattice}
\eeq
By substituting $(x_{hj}, y_{hj})$ couples in Eq. \ref{eq:AF}, the far-field factor reads
\beq
F_N(X,Y)=\frac1N\frac{\sin(\frac{\sqrt{N} X}{\sqrt{N}-1})}{\sin (\frac{X}{\sqrt{N}-1})}\frac{\sin(\frac{\sqrt{N} Y}{\sqrt{N}-1})}{\sin (\frac{Y}{\sqrt{N}-1})} 
\label{eq:latticeAF}
\eeq
whose first minimum is given at
\beq
X_*=\frac{\sqrt{N}-1}{\sqrt{N}}\pi
\eeq
where we defined the adimensional variables $X=k x d/z$ and $Y=k y d/z$. In the large $N$ limit, Eq. (\ref{eq:latticeAF}) becomes

\beq
\begin{aligned}
\lim_{N\rightarrow\infty}
F_N(X,Y)&=\frac{\sin X}{X}\frac{\sin Y}{Y}, \qquad
X_*\rightarrow\pi&
\label{eq:lattice}
\end{aligned}
\eeq

We note that the width of the main lobe in the large $N$ limit becomes independent of $N$, but still varies according to the array size. Specifically, for a given $N$, what it is important for the divergence is the OPA side length. $|F_N(X,0)|^2$ is shown in Fig. \ref{qc:FNl} in the positive $X$ range, for different emitter numbers.

\subsection{Outline of the optical transmitter system}

Here we aim at assessing the scale of the transmission system in terms of the OPA size by combining the previous results on the far-field pattern with characteristics and requirements  of the  ToL system. The main constraints  are the small dimension of GCs and the  size of the OPA that may be realized on the sail, considering also the necessity of the  light splitting, branches harnessing, and phase modulation for the OPA functioning. 

About the first issue, the waist size of the single source is typically of tens to hundreds of micrometers. However, this may be extended to the millimeter range by using metalenses or particular designs \cite{Mar}. 
About the second point, on the other hand involves the minimum FWHM, set by the OPA arrangement, and the maximum intensity gain, arising from the coupler constructive interference.

 We consider as target angular divergence of the main lobe $1$ \si{\micro\radian} which corresponds to less than an astronomical unit at the Earth position, as justified above. 

The main lobe of the lattice configuration for equally spaced GCs is bounded by $X_{min}= kx_{min}d/z = \pi$, where $2d$ is the array side length. 
For sufficiently large $N$, we showed that the array main lobe divergence $\theta_m$ depends critically on the parameter $d$, and in particular this decreases for increasing $d$. 


Since $\theta= x/z$ 
, the array side can be chosen to achieve the desired  divergence according to:

\beq
\theta_m=\frac{x_{min}}{z} \sim 1 \hskip 0.1 truecm  \mu rad  \qquad\mathrm{and} \qquad d=\frac{\lambda}{2\theta_{m} }
\eeq

By using Eq. (\ref{eq:qc_central_int}), we recall that the peak intensity on axis is proportional to $N$ even if the array is spaced widely. 
It is known that in this case the main lobe  shrinks and an increasing fraction of the power is distributed in side lobes. Noticeably, this  effect is resulting from the OPA diffraction and is generally detrimental, as it's diverting part of the emitted power away from the axis. This quantity is quantified by the {\sl thinned array curse theorem} as the relation of the central lobe  power that is reduced by the ratio of the filled area to the empty area \cite{Forw}. 
The power carried by the resulting central lobe is then reduced by the factor 

\beq
\frac{1}{d_{sep}^2 - d_{GC}^2}
\eeq

where $d_{sep}$ is the pitch and $d_{GC}$ is the side of the  GC, considered as square.


To assess the performances of the OPA in three realistic cases, three different couplers designs are considered: a GC coupling a mode of diameter $\sim30$ \si{\micro\meter}, one with diameter $1$ \si{\milli\meter} and, finally, a layered structure composed by the GC surmounted by a metalens, which is spaced in order to produce a $\sim 1$ \si{\centi\meter} diameter output mode. 
The number of GC required to achieve the desired divergence and an $A_{eff}^{TX} = 2.5$ \si{\centi\meter}$^2$ are shown in Tab. \ref{qc:tab}. In the second column it is shown the maximum steering angle $\theta_{Max}$ of the array, which is set at the Gaussian envelope width of the individual emitter, the single GC. Indeed,  the steering phase  added to the OPA is causing the rotation of the lobe and is modulated by the amplitude of the individual emitter, as resulting from Eq. \ref{eq:G} and discussed below in more detail.

 \begin{table}[h!]
 \centering
 \caption{Estimation of array cardinality $N$ made with GCs with waist $w_0$ giving  a central lobe divergence of $1$ \si{\micro\radian} with $A_{eff}^{TX} \simeq 2.5$ $cm^2$}
\begin{tabular}{m{0.1\textwidth} m{0.12\textwidth} m{0.15\textwidth}}
    \toprule
    $w_0$ [\si{\micro\meter}] &  $N$  & $\theta_{Max}$ [\si{\milli\radian}] \\
    
\midrule
30  & $2.8\times10^5$ & 8.50 \\
500  & 1000 & 0.51 \\
5000  & 10 & 0.05 \\
\bottomrule
\end{tabular}
 \label{qc:tab}
\end{table}

\subsection{Fine pointing optimization with beam steering}

As long as we are interested on the  peak intensity  from a transmitter total active area $A_{eff}^{Tx}$ of an array with cardinality $N$, the coupling  of the $N$  waists $w_0$ is  equivalent to the use of a single aperture with waist $\sqrt{N}w_0$. Nevertheless, one of the main advantage of using an array of coherent emitters is the capability of steering the main beam lobe.

The beam steering techniques can be performed by applying a different phase $\phi_j$ to each GC. In this case, the function 
$F_N(\x)$ becomes:
\beq
F_N(\x)=\frac1N\sum_{j=0}^{N-1}e^{i\phi_j}e^{\frac{i k}{z}\x\cdot \x_{0j}}	\,,\qquad F_N(\x=0)=\frac1N\sum_{j=0}^{N-1}e^{i\phi_j}
\eeq

For the sake of simplicity, let's consider a tilt of the beam along the $x$ axis. With this in mind, we need to apply a phase that is proportional to the beam centers $x_{0j}$, namely $\phi_j=-\alpha k x_{0j}$. This means that the far-field beam is shifted on $x$ axis at the receiver by $\alpha z$, namely it is tilted by an angle $\alpha$ and $F_N(\x)$ reads
\beq
F^{\rm steer}_N(\x)=\frac1N\sum_{j=0}^{N-1}e^{-i\alpha k x_{0j}}e^{\frac{i k}{z}\x\cdot \x_{0j}}
=\frac1N\sum_{j=0}^{N-1}e^{i \frac{k}{z}[(x-\alpha z)x_{0j}+yy_{0j}]}
\eeq

By recalling that the overall far-field intensity pattern is given by Eq. (\ref{eq:ffpatterntot}), the Gaussian envelope causes the main lobe of the steered beam to be in general smaller than the on-axis value\cite{Wolfgang1994}. This effect set ultimately a limit to the useful steering angle, that can be considered as a fraction of the indivudual GC divergence. 

Moreover, the OPA lobe angular  width  sets  the minimum $\alpha$ that is needed to be applied. To clarify this, let's consider a lattice array including a sufficiently large number of emitters and it's far-field pattern on $x$ axis. The main lobe width is limited by $2X_*$, where $X_*$ is expressed by Eq. (\ref{eq:lattice}).
\beq
2X_*=\frac{k2dx}{z} \sim 2\pi
\eeq
The corresponding angular aperture is therefore
\beq
\frac{x}{z} \sim \frac{2\pi}{k2d}=\frac{\lambda}{2d}
\eeq

\begin{figure}[H] 
\centering\includegraphics[width=13cm]{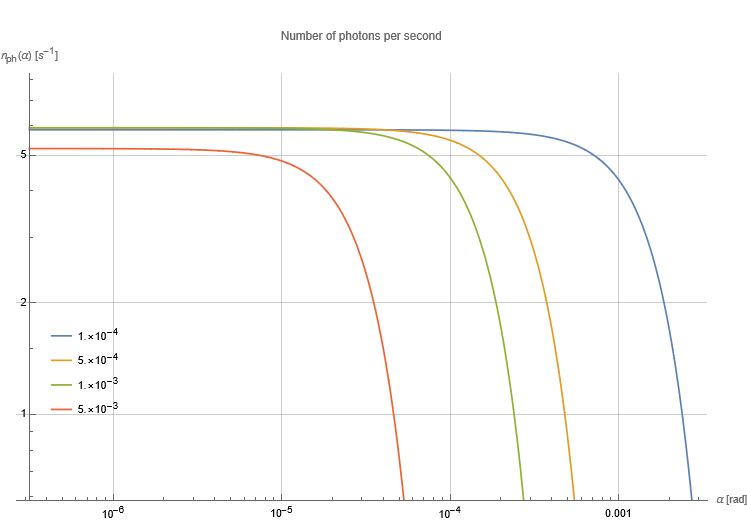}
\caption{Received number of signal photons per second at $\lambda = 800$\,nm and receiver effective area $1$ km$^2$, as a function of the steering angle, for different waists (expressed in the legend in meters) and N apertures. Specifically, to have  $A_{eff}^{TX}\sim 2.5$ \si{\centi\meter}$^2$ for $w_0 = \{100,500,1000,5000\}$ \si{\micro\meter}, we set $\sqrt{N}=\{159, 32, 16, 3\}$.}
\label{fig:numPhTilt1}
\end{figure}

Given that $\lambda/\pi w_0$ is the angular aperture of the individual GC Gaussian envelope, taken where the intensity decays of a factor $1/e^2$ with respect to the peak, then $\alpha$ can be bounded in the range
\beq
\frac{\lambda}{2d} \leq \alpha \leq \frac{\lambda}{\pi w_0}.
\label{eq:range_steering}
\eeq
Lower values of $\alpha$ are not significant being within the main lobe and larger values are imposing a stron attenuation to the resulting tilted lobe.
The interpretation of this latter effect can be understood since as the phase shifts on the array is deviating the lobe direction, this latter is also deviating with respect to the emission of the individual GCs, that remain oriented as the normal to the sail surface. This effect is assessed in  Fig.~\ref{fig:numPhTilt1}, where we plot the number of photons received as a function of $\alpha$, for different waists of the single Gaussian mode, number of GCs in the OPA and a $2d=0.4$ \si{\meter}. The number of emitters is chosen so as to have $A_{eff}^{TX}\sim 2.5$ \si{\centi\meter}$^2$.

The number of angular deviations that are significantly different, that is the lobe is not significantly overlapping, may be approximated by
\beq
n_{\alpha} \sim \frac{2d}{\pi w_0}
\eeq
In principle, we can estimate the number of bits in which the signals delivered to the modulators have to be encoded, i.e.
\beq
n_{b} = \log_2 n_{\alpha} \sim \log_2 \left(\frac{d}{\pi w_0}\right) + 1
\eeq
as well as the phases $\phi_j$ which must be provided by the modulators. Specifically, to cover the full angular interval, the maximum phase difference of the optical signals corresponding to the locations $x_{0j}= \pm \sqrt{2}d$, on the diagonal of the lattice, has to be
\beq
\Delta \phi^{Max} = k\alpha^{Max} 2\sqrt{2}d = k\frac{\lambda}{\pi w_0} 2 \sqrt{2}d = 2\sqrt{2}\pi n_{\alpha}
\eeq

With regard to adjacent emitters, their angular separation vanishes for large $N$ and the same is true for their phase difference.

This aspect is posing the problem of the resolution of the phase modulators, both in terms of the minimum and maximum phase shift to introduce and the corresponding dynamical range of the control system. 

The modulation can be realized by using integrated devices along the branches of the ToL. Due to the limitation in power, mass and the available complexity, and the need to control the phase accurately for many different GCs, a solution exploiting integrated and high-efficiency electro-optics material will be here described.

With reference with state-of-the-art deposition technique and considering lithium niobate as a electro optic material \cite{Zhang21,He19}, the $V_\pi cm$ product may be realized as low as 1.5\,Vcm and with a fully integrated design along a single mode waveguide.
The slightly lower value of 1.4 Vcm was reported for silicon-on-insulator waveguides of thickness of 250\,nm \cite{Feng10}.
Despite a nominal estimation based on modulator lengths, the voltage levels used for phase modulation require to be separately calibrated depending on the particular implementation. Random variations in waveguide, modulator and GC dimensions, as well as manufacturing imperfections, prevent en fact a theoretical assessment of the far-field distribution response to the applied voltages. The minimum number of calibration are $N N_{DAC}$, where $N_{DAC}$ are the number of available phase modulation levels\cite{Guerber2022}.
Moreover, due to the random nature of the calibration, each modulator has to be designed to cover the entire $[0,$ $2\pi]$ range so as to ensure the sought OPA response in all circumstances.

\subsubsection{On the polarization of the optical signal}

The OPA model described above is independent from the  state of polarization of the emitted radiation. At the same time, the polarisation modulation may be exploited for the sail identification and  the coding synchronization, in the perspective of using a  PPM code as described below. The use of this degree of freedom of the photon is crucial in the quantum communications along space channels \cite{Vallone15, Liao17}.

To exploit this possibility, the  capability of polarization modulation at the transmitter and a polarization sensitive receiver are then needed. About the most convenient polarization state,  we may consider that a circular polarization make the link insensitive to the alignment of the sail with respect to the receiver in the rotation around the propagation axis.  Moreover, current  high sensitivity detector as the superconducting nanowire single photon detectors (SNSPD) with a meander as the anode,  have a maximum for a particular  input linear  polarization  state. 
The circular polarization can be then exploited for  transforming  the  collected photons in that optimal  linear polarization state by a birefringent plate close to the receiver focalplane. This is a realized with a quarter-waveplate plate with the  22.5° orientation of the optical axis with respect to the main meander axis. 

The study reported in the final section describe how to generate the circular polarization state at the transmitter. 
We note that the possibility to actively control the polarization behaviour of the metalenses, which is a possible evolution of the current technology, may be envisaged to obtain a polarization modulation of the single GC scheme discussed in this work. 


From these considerations, it follows that most of the efforts has to be focused on the pointing capability, which comprises both a precise evaluation of the sailcraft-Earth relative position and a consistent steering of the Far-field distribution of the transmitted light. The latter implies that the correct calibration of the phase modulators is also a fundamental and delicate task, as well as the evaluation and eventual correction of any deformations in the sail that could affect the OPA spatial distribution.

\newpage
\section{High coupling efficiency  transmitting leaves}\label{sec:leaves}

In this  section we discuss on the optimization of the emission efficiency from the individual grating coupler and on the feature to impose a circular polarization to overcome the alignment requirements with receiver system as well as the flattening of the final wavefront. 

The value of the wavelength suitable forr the Starshot transmitter is currently still not fixed. Indeed it will result from the optimization of several factors including the diffraction width of beam, directly proportional to $\lambda$, the technology of the laser source, the optical harnessing and the modulations and of the ToL leaves. In this section we adopted the value of 800 nm, that represent a candidate value that optimize the generation section.

\subsection{High efficiency apodized grating couplers}
A solution to couple the light coming from a waveguide with high efficiency into the vertical direction is given by Binary Blazed Grating Couplers (BBGCs). They exhibit high transmission efficiency but have a limitation in terms of outgoing beam waist (5-10 $\micro$m). \cite{xu} To overcome this limitation we design different optics able to generate high efficiency coupling with larger beam waist which are called apodized grating couplers (AGC). The apodized grating couplers are grating with a variation in terms of fin's width along the whole structure; this approach permits both to design  the grating in order to enhance the coupling efficiency into a target mode and to increase the outgoing beam waist. \cite{taillaert} - \cite{marchetti}
AGCs have some limitations in terms of perfect vertical directionality for the coupled mode, but, this problem can be overcome by using a phase corrector after the grating in order to correct the deviation from the vertical direction; furthermore, the phase corrector can act also as collimator.

\begin{figure}[h!]
\begin{center}
\includegraphics[width=11 cm]{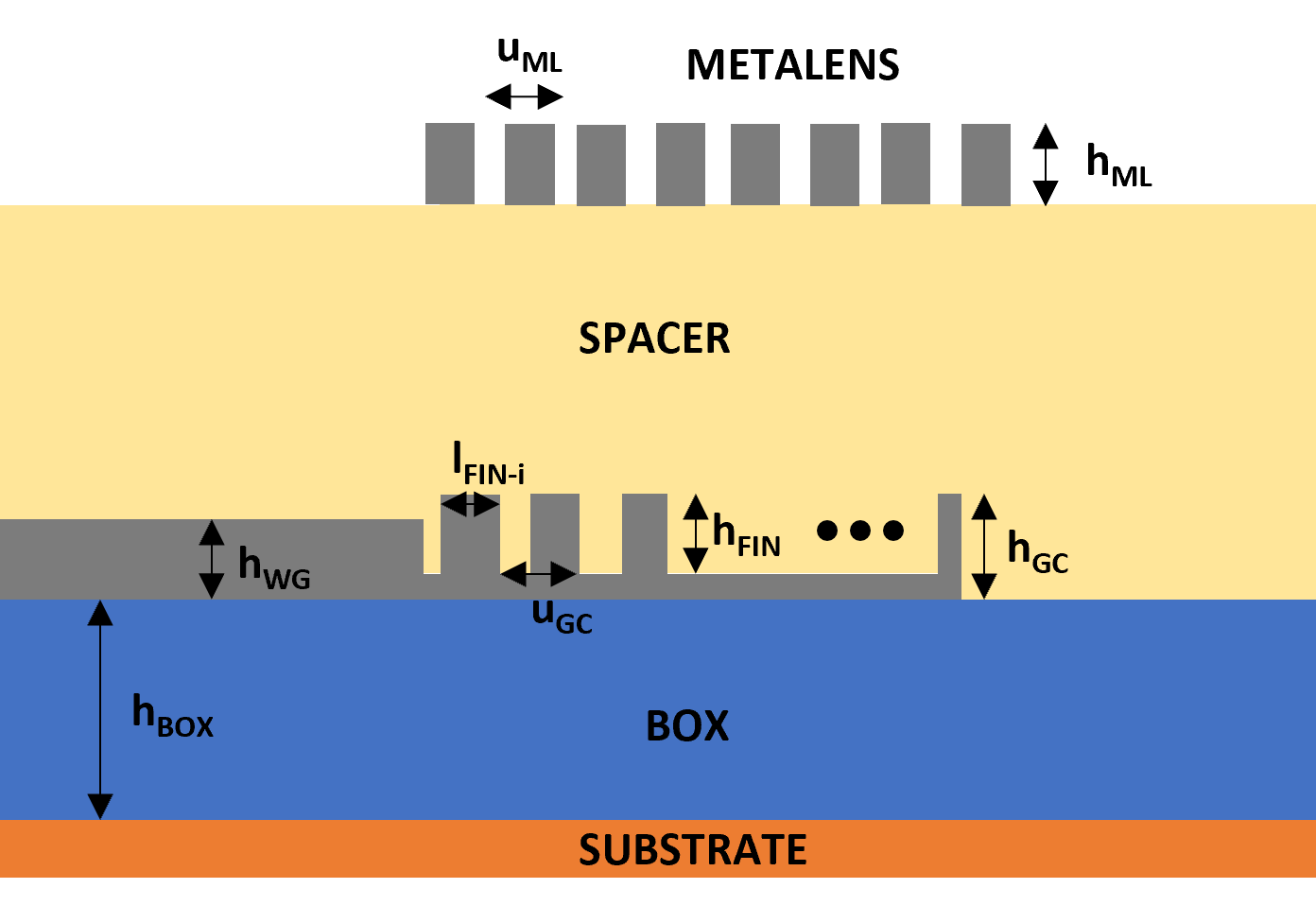}
\caption{Schematic configuration of the grating coupler and phase corrector.}\label{fig:noGCML}
\end{center}
\end{figure}

For our aims we proposed AGCs, as shown in Figure \ref{fig:noGCML}, made of silicon nanofins having different length ($l_{FIN-i}$) and same height ($h_{FIN}$), placed into a digitalized pattern with period $u$. The AGC pattern is placed onto a  substrate, which can be considered of silicon, whose total height is $h_{GC}$ and the light is provided by a silicon waveguide of height $h_{WG}$. The above-described components are fabricated over both a BOX layer ($h_{BOX}$) to ensure a good refractive index contrast and a substrate ($h_{SUB}$) to attach the structure onto the sail. The structure is covered by a material, called spacer, which acts as substrate for the phase corrector modelled as a metalens.
Since the goal of the leaves is to couple as much as possible the light incoming from the waveguides to a collimated gaussian mode outgoing the leaves themselves,  we simulated the behaviour of different AGCs changing some parameters in order to maximize the coupling efficiency (${CE}$).
We estimated the coupling efficiency as:
\begin{equation}
\label{eq:CE}
CE = T \cdot OI
\end{equation}
being $T$ the transmission of the outgoing mode calculated as \begin{math}T=abs(E_{OUT})^2/abs(E_{IN})^2\end{math}, and $OI$ the overlap integral between the simulated coupled mode and a gaussian mode:
\begin{equation}
\label{eq:OI}
OI = \frac{|\iint_S E_1 (x,y) \cdot E_2^* (x,y)\,dx \,dy|^2}{\iint_S |E_1 (x,y)|^2 \,dx \,dy \cdot \iint_S |E_2 (x,y)|^2 \,dx \,dy|^2}
\end{equation}
where $E_2(x,y)$ is a gaussian mode and $E_1x,y)$ is the grating coupler's outgoing mode.

We started optimizing the transmission at $\lambda = 800nm$ of the coupled mode by varying the parameters such as the height of the waveguide, the grating coupler's height etc. and fixing the refractive indexes of the materials (Table \ref{tab:parametersGC}) It has been extrapolated that the maximum transmission ($T_{800nm}=0.8135$) is obtained when the parameters take the values depicted in Table \ref{tab:parametersGC}.
\begin{table}[h]
\centering
\begin{tabular}{l | l | l | l | l | l | l | l | l}
$u$ [\micro m] &  $h_{WG}$ [\micro m] & $h_{GC}$ [\micro m] & $h_{FIN}$ [\micro m] & $n_{Si}$ & $h_{BOX}$ [\micro m] & $n_{BOX}$ & $n_{SUB}$ & $n_{SPACER}$\\
\hline
0.280 & 0.160 & 0.210 & 0.155 & 3.5 & 1.2. & 1.5 & 3.5 & 1.5 
\end{tabular} 
\caption{Optimized parameters values for AGC simulations in order to maximize the transmission ($T_{800nm}=0.81$) of the outgoing mode}
\label{tab:parametersGC}
\end{table} 
\begin{figure}[h!]
\begin{center}
\includegraphics[width=11 cm]{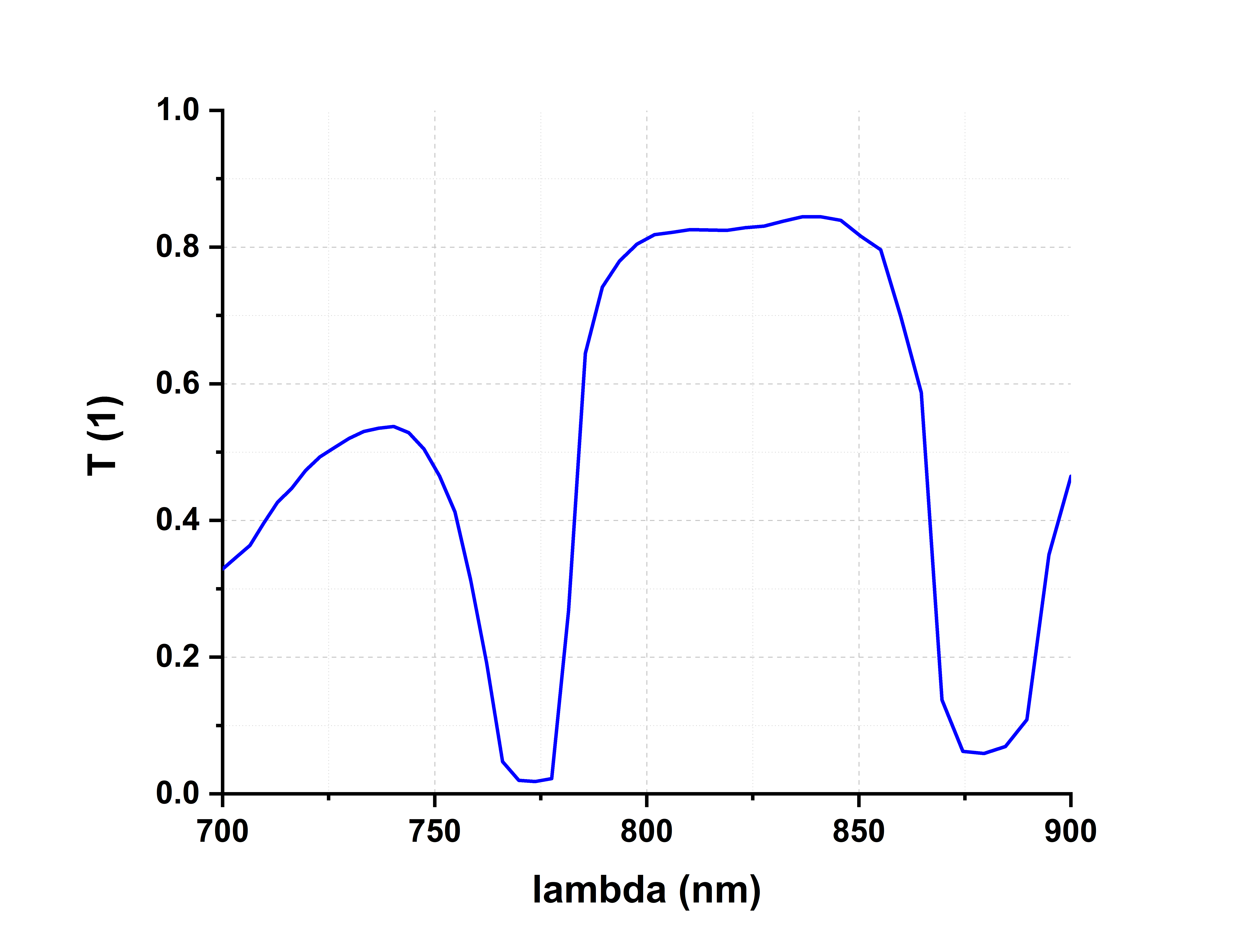}
\caption{Transmission values at different wavelenghts of a coupled mode using a periodic grating coupler following the parameters depicted in Table \ref{tab:parametersGC}.}\label{fig:noT}
\end{center}
\end{figure}

After that, we simulated the behaviour of our GC apodizing the duty cycle along the structure; the duty-cycle has been defined as $dc = l_{FIN}/u$ and the apodizing process ensure that the duty-cycle is different for each GC's nanofin, so we have $dc_i = l_{FIN-i}/u$. This apodization in necessary in order to vary the coupling strength ($\alpha$) of a grating coupler along its whole length. The coupling strength of a grating coupler, as mentioned in \cite{taillaert}, is the constant of power's exponentially decay along a grating coupler with uniform duty-cycle (so-called periodical grating coupler). 
\begin{equation}
P=P_0 \cdot exp(-2 \alpha z)
\label{eq:P}
\end{equation}
which, in the particular case of the apodized grating coupler, becomes a function of the position $z$ ($\alpha(z)$) along the GC's length.
As proposed in \cite{marchetti}, we decided to use a linear apodization based on the linear variation of the grating coupler's duty cycle along its entire length; the formula for the apodization can be written as:
\begin{equation}
\label{eq:dc}
dc(z) = dc_0-kz
\end{equation}
being $dc_0$ the is the initial duty-cycle of the first nanofin, $k$ is the linear apodization factor and z is the distance of each nanofin from the starting point of the grating.
We simulated the behaviours of an AGC made of 250 periods, so $70\micro m$ length, with a initial duty-cycle of $dc_0 = 0.975$ and a linear apodization factor $k = 0.6$, and all the others parameters following Table \ref{tab:parametersGC}.

\begin{figure}[h!]
\begin{center}
\includegraphics[width=11 cm]{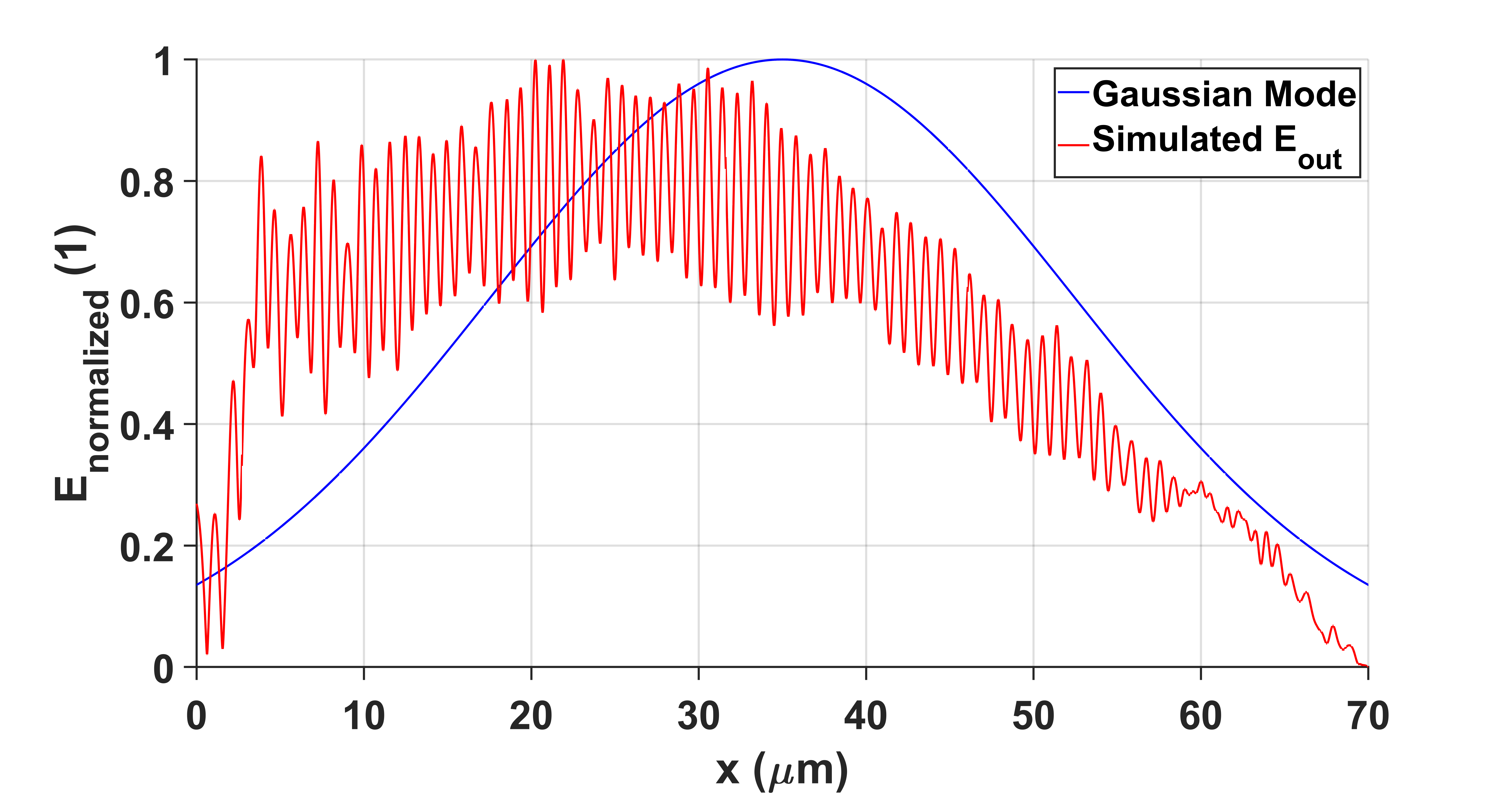}
\caption{Simulation of the coupled mode using an AGC made of 250 periods, so $70\micro m$ length, with a initial duty-cycle of $dc_0 = 0.975$ and a linear apodization factor $k = 0.6$ and all the others parameter following Table \ref{tab:parametersGC}.}\label{fig:noCoupling}
\end{center}
\end{figure}

As shown in Figure \ref{fig:noCoupling} the coupled mode has a quasi-gaussian shape and calculating the overlap integral with a gaussian beam (having a beam waist $w_0 = 70\micro m$) using Eq. \ref{eq:OI} it obtains a value of $OI = 0.8783$. Combining both the overlap integral and the transmission it is possible to calculate the coupling efficiency of our proposed APG; substituting the extrapolated values into Eq. \ref{eq:CE} it obtains a coupling efficiency of $CE=0.8135\cdot0.8783=0.7145$

\begin{figure}[h!]
\begin{center}
\includegraphics[width=11 cm]{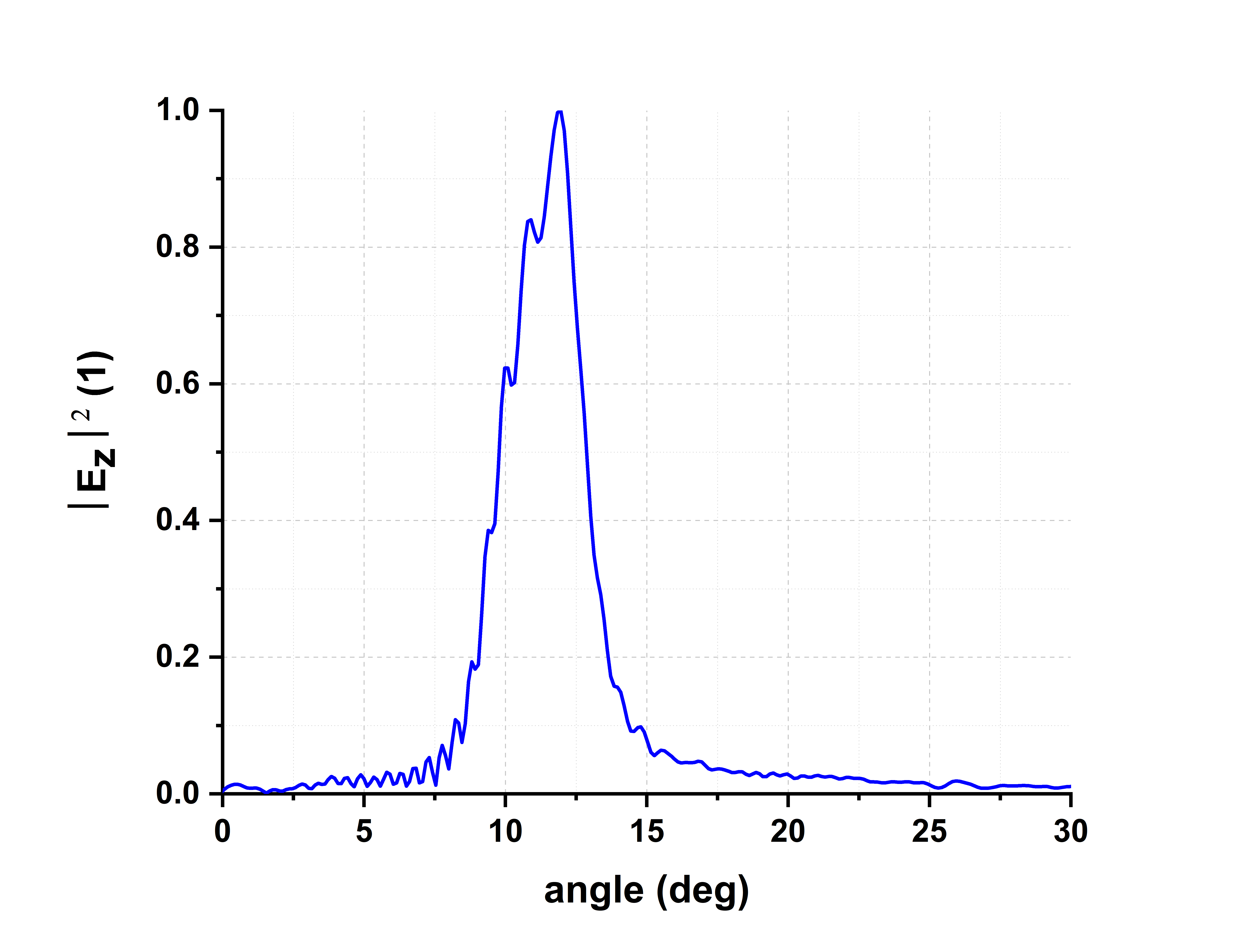}
\caption{Far-Field intensity of the coupled mode using an AGC made of 250 periods, so $70\micro m$ length, with a initial duty-cycle of $dc_0 = 0.975$ and a linear apodization factor $k = 0.6$ and all the others parameter following Table \ref{tab:parametersGC}.}\label{fig:noFF}
\end{center}
\end{figure}

\begin{figure}[h!]
\begin{center}
\includegraphics[width=11 cm]{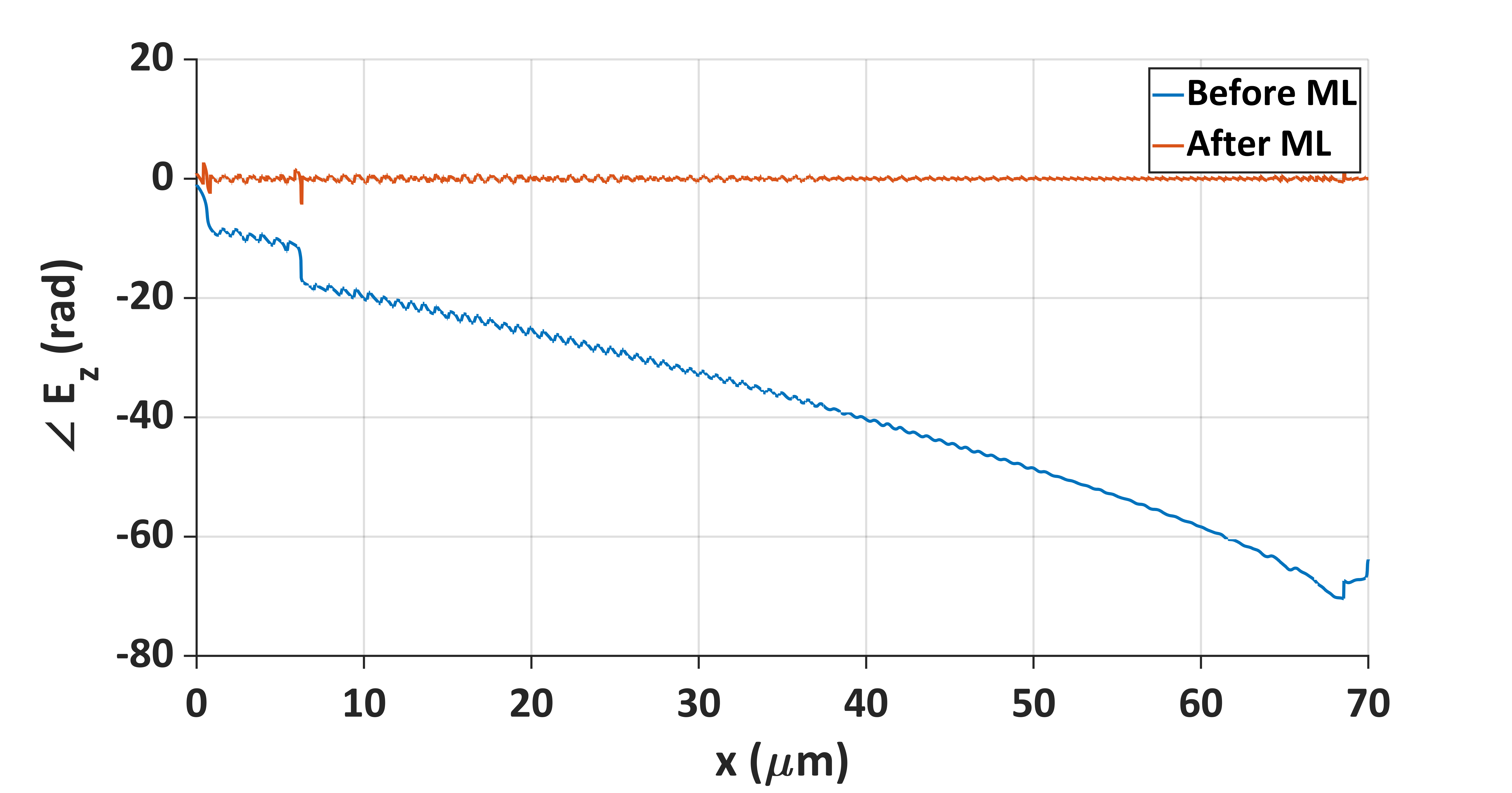}
\caption{With a blue line it is represented the simulation of the coupled mode's phase at $y=8\micro m$ using an AGC made of 250 periods, so $70\micro m$ length, with a initial duty-cycle of $dc_0 = 0.975$ and a linear apodization factor $k = 0.6$ and all the others parameter following Table \ref{tab:parametersGC}. With the red line it is represented the correction performed by the metalens.}\label{fig:noFFang}
\end{center}
\end{figure}

Furthermore, we evaluated the directionality of the coupled mode which isn't perfectly vertical (as previously discussed). Figure \ref{fig:noFF} shows that electric field propagation wavevector is inclined by 12 degrees from the grating coupler's orthogonal direction and carries both a linear phase gradient (Figure \ref{fig:noFFang}) and a slight curvature which must be corrected using a metalens acting as phase corrector.

From the results described above, the proposed solution using the apodized grating couplers offers an improvement in terms of both coupling efficiency and size of the outgoing mode rather than the binary blazed grating approach. In fact, as depicted above we reached a Coupling Efficiency higher than 0.7 and a quasi-gaussian outgoing mode with a beam waist of $70\micro m$, at the contrary, with the bineary blazed grating approach it can be reached quite similar CE but the outgoing beam waist is only few micrometers. The main limitation is the directionality of the beam but this problem is overcome using a phase corrector.

\subsection{Metalenses for phase correction and polarization conversion}
As described in the previous section, the mode coupled using the apodized grating coupler is tilted from the orthogonal direction of a certain quantity (i.e., 12 degrees) (Figure \ref{fig:noFF}), so it carries a linear phase gradient during the propagation combined with its natural divergence (Figure \ref{fig:noFFang}). This situation requires a phase correction element in order to redirect the directionality of the outgoing beam in the orthogonal direction and without imposed divergence. It is possible to integrate theses functionalities using diffractive optics that works well but can only correct the phase, so we considered to turn into a metasurfaces approach which permits to encode more functionalities rather than the only phase correction. In particular, we decided to implement the phase correction using a metalens because whose peculiarity is to be able to encode different functionalities depending on the input polarization state. \cite{Khorasaninejad} - \cite{Zhang} 
One of the feature we would like to encode is the possibility to convert the linearly polarized state of the beam coupled by the AGC into a circularly polarized beam in order to decrease the receivers by a factor 2. To this aim we designed some metalenses able to correct the phase of the coupled mode and , at the same time, to generate a circular polarization state with different spin depending on the input linearly polarized state.
In detail, the metasurface proposed in this work is a dielectric metalens (ML), made of a 2D array of birefringent metaunits that exploit  the dynamic phase. Our DFML is constituted of subwavelength metaunits (MUs), the so-called metaatoms (MAs), arranged over a square lattice, represented by silicon nanopillars on a substrate, surrounded by air. Each pillar belongs to a subset of nanostructures with different cross sections but the same orientation and height, and acts as a quarter-wave plate in order to maximize the polarization conversion and, therefore, the optical efficiency. 

\begin{figure}[h!]
\begin{center}
\includegraphics[width=13.5 cm]{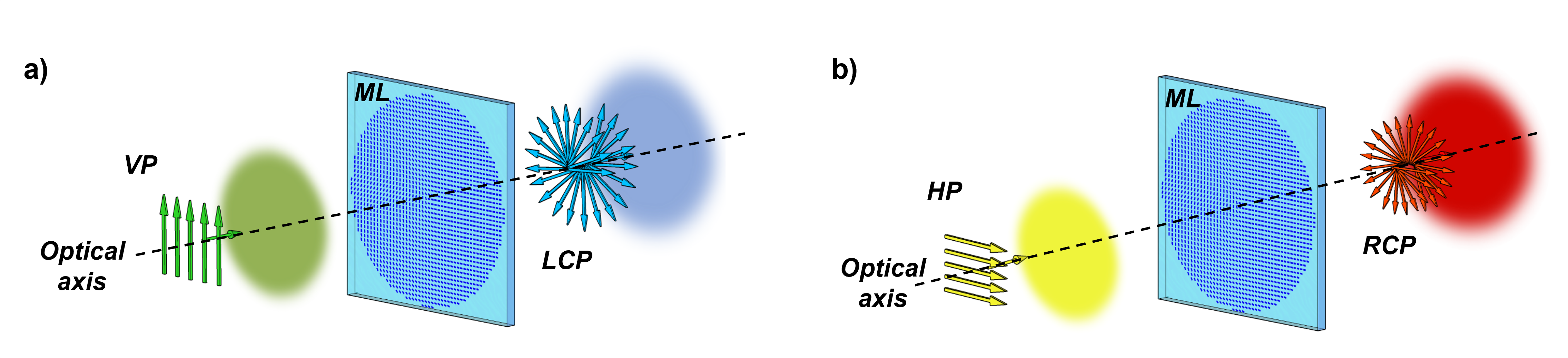}
\caption{Schematic representation of the polarization conversion paradigm described by Eq.\ref{eq:Jv} (a) and Eq. \ref{eq:Jh} (b)}\label{fig:noMLfunc}
\end{center}
\end{figure}

For the benefit of the reader, we provide in the following the theory underlying the working principle of metaunits. In particular, the Jones matrix J for the metaatom at the coordinates (x, y) is:
\begin{equation}
    J=e^{(i \frac{\delta_x + \delta_y}{2})} cos(\frac{\Delta}{2})\begin{bmatrix} 1 & 0\\0 & 1\end{bmatrix}-ie^{(i \frac{\delta_x + \delta_y}{2})}sin(\frac{\Delta}{2})\begin{bmatrix} cos(2\theta) & sin(2\theta)\\sin(2\theta) & -cos(2\theta)\end{bmatrix}
\label{eq:Jones}
\end{equation}
being $\theta$ the local orientation of the metaatom fast axis and $\Delta=\delta_y - \delta_x$ is the phase retardation between the two axes of the metaunit (the spatial dependence has been omitted to simplify the notation).
If it is imposed the conditions $\Delta = \pi/2$ and $\theta = \pi/4$, and recalling the Jones formalism of the polarization states $(|H\rangle=\begin{bsmallmatrix} 1 \\0 \end{bsmallmatrix}, |V\rangle=\begin{bsmallmatrix} 0 \\1 \end{bsmallmatrix}, |L\rangle=\frac{1}{\sqrt{2}}\begin{bsmallmatrix} 1 \\i \end{bsmallmatrix}$ and $|R\rangle=\frac{1}{\sqrt{2}}\begin{bsmallmatrix} 1 \\-i \end{bsmallmatrix})$ it obtains:
\begin{equation}
    J|H\rangle=e^{(i \frac{\delta_x + \delta_y}{2})}|R\rangle
\label{eq:Jh}
\end{equation}
\begin{equation}
    J|V\rangle=e^{(i \frac{\delta_x + \delta_y}{2}+\frac{\pi}{2})}|L\rangle
\label{eq:Jv}
\end{equation}
so properly design each metaatom it is possible both to correct the phase using the dynamic phase contribution ($e^{(i \frac{\delta_x + \delta_y}{2})}$) and to convert the polarization state from linear to circular.(Figure \ref{fig:noMLfunc})

We set up a custom-made Finite-Element Method (FEM) simulation in the wavelength domain (using COMSOL Multiphysics®) to find the best set of metaatoms respecting the ML requirements described above. Each subunit has been defined as a silicon nanopillar ($n_{Si}=3.5$) surrounded by air ($n_{Air}=1$) placed on the top of a substrate ($n_{Sub}=1.5$). All the materials were considered as non-absorptive (n=Re(n), Im(n)=0). Thus, we imposed some conditions to simulate properly the nanostructures: Periodic Port conditions were set in the substrate at a distance equal to   from the nanopillar and at a distance greater than   over the pillar, both to collect the scattering parameters of the structure and simultaneously ensure the far-field regime; Perfectly Matched Layer (PML) conditions have been imposed outside the ports at a distance greater than   to visualize the transmitted and reflected fields, and to absorb the field over a certain distance to avoid unwanted multiple reflections. Finally, Periodic Boundary conditions (PBC) were set (along the xz and yz planes) to permit the correct simulation of the interaction between the various metaunits of the metalens. 

Simulations were performed fixing the period of the metaatoms matrix at $400 nm$ along x-axis and y-axis, and sweeping the sizes of the metaunit cross-section ($L_x$, $L_y$), considering fabrication constraints and the subwavelength regime, at the working wavelength of 800 nm. Moreover, due to fabrication limitations, we imposed a fixed height H of 500 nm. Thus, for a fixed phase delay $\delta_x$ along the fast axis of the pillars, we selected the cross sections satisfying the condition $\Delta = \pi /2$. In particular, we selected metaatoms having a maximum phase difference of 0.05 rad from the QWP condition, i.e. $\Delta_{simulated}-\Delta=0.03 rad$. At the same time, to ensure an homogeneous polarization conversion, we imposed strict conditions on the transmissions for TE and TM polarizations. More precisely, we fixed $|T_{x,i}-T_{y,i}| < 0.05$, being $T_{x,i}$  and $T_{y,i}$ the transmittance of the i-th metaatom for TE and TM polarizations, respectively. 
Concurrently, to guarantee a homogeneous transmittance over the whole metalens we impose a maximum difference of 0.05 in transmittance among the metaatoms $|T_{avg,i}-T_{avg,j}| < 0.05$, $i,j=0,1,...,N$, being $T_{avg,k} = \frac{T_{x,k}+T_{y,k}}{2}$ .

\begin{figure}[h!]
\begin{center}
\includegraphics[width=13.5 cm]{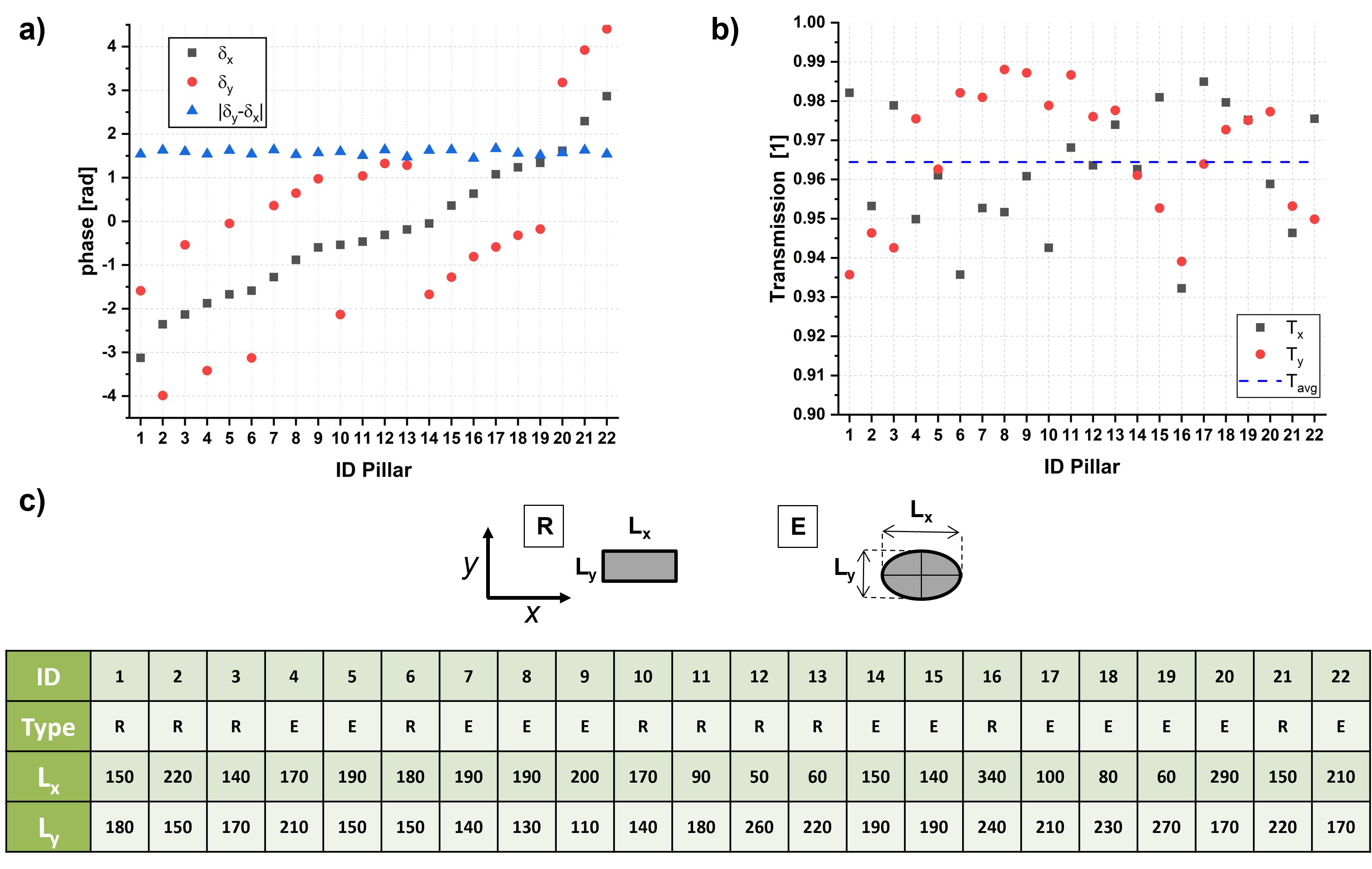}
\caption{Library of different silicon nanopillars working at $800nm$ providing the recipe to built-up a dual-functional metalens. (a) Phase delays for TM (x-delay) and TE (y-delay) polarizations, (b) Transmittance of each nanopillar under TM ($T_x$) and TE ($T_y$) polarization compared with the average transmittance ($T_avg$) calculated among all the transmittance values. (c) Different types of pillars composing the meta-library. Rectangular and elliptical pillars (R-E). Below, the complete list of the metaatom library showing the type of pillars and the corresponding size.}\label{fig:noMA}
\end{center}
\end{figure}

As a consequence, the previous requirements limit significantly the choice of possible cross-sections for the given thickness and shape. Therefore, in order to increase the degrees of freedom to find the adequate set of nanostructures covering the whole 2$\pi$ range, different shapes have been considered, such as rectangular and elliptical. A meta-library of 22 different nanopillars has been extrapolated from the simulations, which permits to have a well distributed 22-level discretization of the phase over the range 0-2$\pi$ (Figure \ref{fig:noMA}).
Figure \ref{fig:noMA}a shows that the range 0-2$\pi$ has been covered very well, and depicts two different types of behaviour among the pillars, with both $\delta_x > \delta_y$ as well as the opposite. 

These configurations always respect the QWP restriction but differ in terms of pillar's rotation, in fact, when a pillar shows $\delta_x < \delta_y$ we must rotate it of a quantity $\theta = \pi/4$, on the other hand when we have $\delta_x > \delta_y$ the nanopillar must be rotated of an angle $\theta = -\pi/4$ (note: all these configuration can be derived from Eq.\ref{eq:Jones}).
Moreover, Figure \ref{fig:noMA}b depicts the transmission of all the selected configuration under different input polarization state. It is worth noting that our metasurface reaches an average transmission higher than 0.965.

Finally, We implemented to our metalenses a phase correction in order both to collimate and redirect the impinging beam coming from the GC. As depicted in Figure \ref{fig:noFFang} with a red line, the curvature of the field is absent so the metacorrector works as espected.
In conclusion, the proposed metasurface is able both to correct the linear phase gradient and the curvature of the AGC's coupled mode acting on the dynamic phase variation along the whole nanostruture. At the same time, it is able to convert a linearly polarized light into a circularly polarized one whose spin depend on direction of the impinging linear polarization. So, accurately design the metalens and choosing the correct metaatoms along the entire structure depending on the required phase to be corrected we are able to set the coupled mode both collimated and directed along the orthogonal direction of the sail with an efficiency higher than 0.965.


The feasibility of coupling the light coming from a waveguide into a quasi-gaussian mode propagating in free-space was  discuss in the previous sections, reaching  the ability both to correct the phase distorsions and to convert the polarization state of the coupled light by means of metasurfaces.
We depict a configuration able to generate a circularly polarized gaussian beam with a $70\micro m$ waist and a coupling efficiency higher than 0.7. 

We note that this design paradigm can be used  to obtain also larger beam waist. As a matter of fact, by properly designing the size, the divergency of the apodized grating coupler and the distance between the AGC and the metacorrector, it is possible to adjust the waist of the gaussian beam keeping the same coupling efficiency.



\clearpage
\section{Identification of a sails in the flight of sails}
\label{sec:sailident}

In the vision of the Starshot Project, it is considered that not one but  that a fleet of $N_sail$ sails will be launched with a suitable periodicity, possibily addressing the measurement of different observables. In order to isolate the individual messages,  a method for the identification of the individual sailcrafts should be designed. 
For this purpose, we propose to exploit the actual Doppler shift determined by the acceleration phase, where slight differences in the final speed can cause large frequency shifts, with respect to the nominal Doppler shift.


The $N_sail$ sails are accelerated by means of a  photon engine on the Earth, providing the same acceleration, on average. However, some random differences may occur, and the sails actual speeds might differ, giving rise to different frequency (Doppler) shifts. 
Our aim is to exploit these frequency shifts to identify the $N_sail$ sails. 
It is worth noting that in this scenario the \emph{relativistic} Doppler shift should be considered. Therefore, the Doppler shift $f_d$ for a sail with speed $v$ is 
\begin{equation}
   f_d = \left(\frac{ \sqrt{1-v/c} }{ \sqrt{1+v/c} } - 1\right)f_0\label{eq:doppler}
\end{equation}
with $f_0$ the received carrier frequency.
Then $f_d$ is no longer linearly related to $v$. 
However, if the differences between sail speeds are small compared to $c$, relation  (\ref{eq:doppler}) can be linearized. 

We assume that the $N_sail$ sails speeds are modeled by $N_sail$ independent Gaussian random variables with variance $\sigma_v^2$. In a linearized  model the standard deviation of the Doppler shift $\sigma_f$ is  $\sigma_f = \sigma_v/c f_0 $. 
The condition for the separability of the $N_sail$ sails is that their frequency shifts are separated more than the bandwidth $B$ of the optical filter used at the receiver. 
Considering the same carrier frequency $f_0$ for all the $N_sail$ sails, if $f_i$ is the actual shifted frequency of the sail $i$, the relationship $| f_i - f_j | > B$,  $\forall i\neq j$, must hold in order to distinguish $i$ from $j$.
\begin{figure}[h!]
\begin{center}
\includegraphics[width=0.60\textwidth]{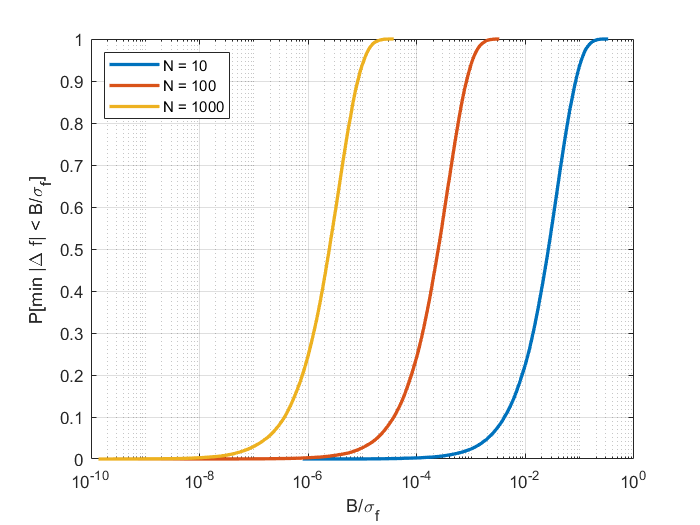}
\caption{ $P [\min | f_i - f_j| < B/ \sigma_f ]$ as a function of $B/ \sigma_f$ for $N_sail=\{ 10; 100; 1000 \}$ sails.}\label{fig:dopp2_gauss}
\end{center}
\end{figure}

The results shown in Fig.~\ref{fig:dopp2_gauss} allow us to evaluate the likelihood of non-overlapping transmissions from distinct sails.
Specifically, Fig. \ref{fig:dopp2_gauss} shows the probability that $\min \{ | f_i - f_j|, \forall i,j=1,\ldots,N_sail; i\neq j \}$, is smaller than $B$.

\begin{figure}[h]
\begin{center}
\includegraphics[width=0.62\textwidth]{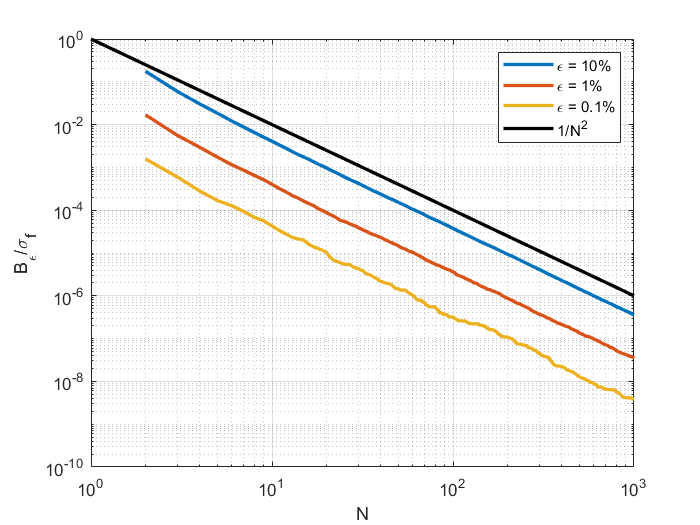}
\caption{Maximum $B/ \sigma_f$ varying the number of sails $N_sail$, for different likelihoods $\epsilon = \{ 0.1; 0.01; 0.001\} $.}\label{fig:dopp1_gauss}
\end{center}
\end{figure}

\begin{figure}[hb!]
\begin{center}
\includegraphics[width=0.63\textwidth]{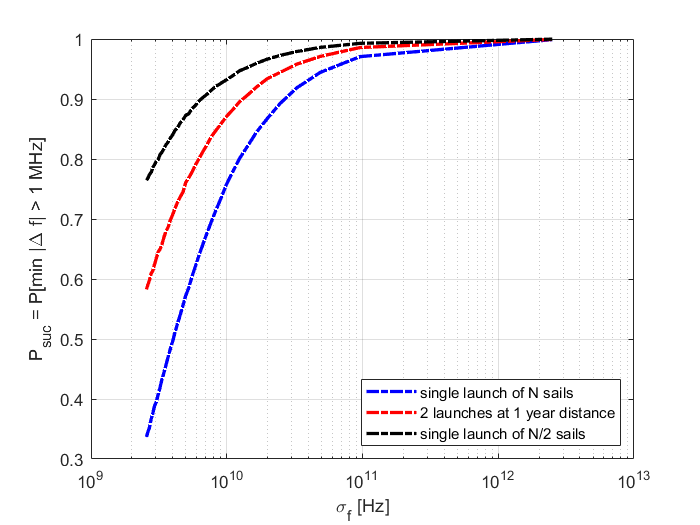}
\caption{ $P [\min | f_i - f_j| > B ]$ considering $N_sail=100$ sails, with clusters at 1 year distance and $B=1$\,MHz.}\label{fig:diff_gauss1}
\end{center}
\end{figure}

Fig. \ref{fig:dopp1_gauss} shows the maximum  $B/ \sigma_f$ needed to guarantee a probability $\epsilon$ of not distinguishing the sails, as a function of the number of sails $N_sail$.
%
Actually, different launch scenarios could be envisaged: instead of launching the sails all together, they can be launched in clusters at different times. However, to reach $\alpha-$Centauri at the same time, the sails launched later must be accelerated to a higher cruise speed. As a consequence, the Doppler shifts are modelled by Gaussian distributions with a different average value for each cluster.
Fig. \ref{fig:diff_gauss1} shows the probability of distinguishing the sails, with two clusters launched at one year distance. 

\clearpage
\section{On the optimization of the  digital communication system and the channel coding used by the ToL transmitter}
\label{sec:lbdr}

The optical transmission system implemented by the ToL has the objective to send to the receiving station a payload of information bits, including the images and physical measurements obtained at destination, with a total data volume of the order of  Mbits  \cite{Messerschmitt_2020}.
From the analysis of the optical link budget, a very strong attenuation due to the optical diffraction characterizes the communication channel. This fact imposes to envisage  very carefully the scheme for the digital communications that the ToL will realise. This also requires    to design the most suitable configuration of a certain number of parameters, and in particular the best error correction code, fulfilling the constraints imposed by the technology, taking into account the potential changes during the operation time. 

On the base of extensive previous studies devoted to this subject \cite{Messerschmitt_2020}, we consider here that the modulation to be adopted is pulse position modulation (PPM) with a symbol of $M$ slots and slot time $T_s$. The slot time $T_s$ depends on the modulation bandwidth capabilities of the laser at the transmitter, since the optical pulse conveying the position information must be confined within the duration $T_s$. 
A sequence of PPM symbols is organized into a frame of duration $T_F$ for coding purposes. Note that the use of channel coding techniques is mandatory in the photon-starving environment considered. 

\begin{figure}[h]
\begin{center}
\includegraphics[width=0.9\textwidth]{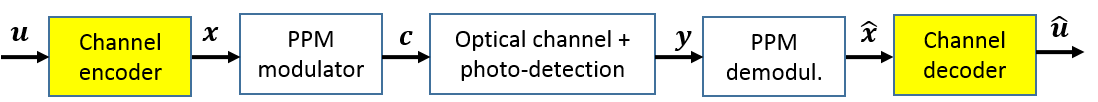}
\caption{PPM system model with channel coding.}\label{coding_model}
\end{center}
\end{figure}

The optical channel shown in Fig.~\ref{coding_model} is a Poisson channel: thus, the received photon arrival is modelled by a Poisson random process, with an average rate of $N_s$ photons per second. The useful signal is corrupted by a background noise, with an average number of photons per second denoted by $n_b$.

Channel coding is performed on a frame-base. With reference to Fig.~\ref{coding_model}, during each PPM frame period $T_F$ a sequence $\mathbf{u}$ of $b_I$ input bits is coded by the error correction code (ECC) into a word $\mathbf{x}$ of length $b_F$, where the ratio $r=b_I/b_F$ is the {\it code-rate}. 
Possibly other $N_{extra}$ bits are added as CRC or termination bits and transmitted at the end of each frame and typically the frame period  $T_F$ includes a guard-time, with a guard time factor $\alpha_{gt}$ (e.g., if the guard time is 25$\%$, $\alpha_{gt}=1.25$).
Then the useful data bit-rate $B_r$ is given by
\begin{equation}
    B_r =\frac{b_I}{T_F} = \frac{b_F \, r - N_{extra}}{M\, T_s\,b_F/\log_2(M)\, \alpha_{gt}}
\end{equation}
For the channel coding, the main system parameter is the code rate $r$. 

In this coded optical communication system, the target requirement is represented by the desired bit error rate (BER). In fact, it can be assumed that the objective image quality can be quantified by a maximum tolerable BER.
On the other hand, the BER in the PPM Poisson channel depends on the mean number of useful signal photons per PPM slot $N_s= n_s  \, T_s$ and on the mean number of noise photons per PPM slot $N_b = n_b  \, T_s$. From the plots of the coded BER, one can infer the minimum number of signal photons $N_{s, min}$, for the required BER.

Note that the useful photon rate $n_s$, as described in detail in Section~\ref{sec:concept}, depends on several other system parameters, such as:
the average transmitted power $P_{tx}$,
 the transmitting and receiving aperture diameters $D_{tx}$ and $D_{rx}$, 
 the overall link losses, including pointing, path loss, etc.

Given all the system parameters and the objective BER, the best channel coding strategy is the one that minimizes the collected photons $N_{s, min}$ to guarantee that BER, and therefore minimizes the time needed to download the payload image.

We remind that the useful signal photon rate (photons per second) $n_s$ can be determined as described in Section~\ref{sec:concept}, while the rate of background noise photons $n_b$ is given by
\begin{equation}
    n_b = \eta_{D} \left[L_{s} (\lambda) \, A_{rx} \Delta \lambda + P_{ext}\right] \,  \frac{\lambda}{hc} + n_{DC}
\end{equation}
where $ n_{DC}$ represents the detector dark count rate, $\eta_{D}$ the detector efficiency, $P_{ext}$ is the residual received laser power without transmission (related to the laser extinction ratio), $\Delta \lambda$ is the spectral width of the receiver filter, while $A_{rx}$ is the effective receiving aperture area, $h$ the Planck constant and $c$ the speed of light. Finally, $L_{s} (\lambda)$ is the total background or stray radiation at the receiver, at the signal wavelength $\lambda$, and not related to the transmitted signal.

\subsection{Channel coding techniques for the Poisson channel - SCPPM}\label{codingSCPPM}

\begin{figure}[h]
\begin{center}
\includegraphics[width=0.9\textwidth]{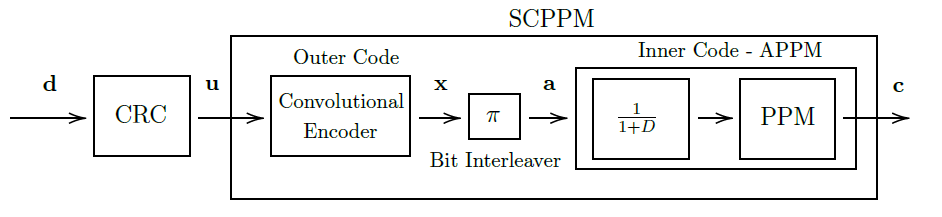}
\caption{Conceptual scheme of the SCPPM encoder.}\label{fig:encoder}
\end{center}
\end{figure}

We first present the {\it serially-concatenated PPM} (SCPPM) scheme which refers to a precise combination of modulation and coding technique mostly used in a deep-space optical link scenario \cite{scppm}. The name SCPPM derives from the combination of a PPM modulator with other blocks serially concatenated to it, namely a convolutional code as the error-control code, an accumulator and an interleaver.
Fig.~\ref{fig:encoder} shows the block diagram of the SCPPM encoder \cite{scppm}. 

The data vector \textbf{u} firstly enters in a cyclic redundancy check (CRC) block, that appends 32 binary digits. The CRC attachment can be used in the receiving phase for the evaluation of the correct decoding of the codeword.

After that, \textbf{u} is convolutionally coded. We consider this encoder, also denoted as \textit{outer code}, as a constraint-length-3 convolutional code with generator polynomial \textbf{g} = [5, 7, 7] in octal notation \cite{scppm}, or
\begin{equation*}
    g^{(1)}(D) = 1+D^2
\end{equation*}
\begin{equation*}
    g^{(2)}(D) = 1+D+D^2
\end{equation*}
\begin{equation}
    g^{(3)}(D) = 1+D+D^2
\end{equation}
generating a 1/3 code rate. It is worth noting that this basic encoder can be punctured in order to achieve also rate 1/2 or 2/3.

Once the codeword $\textbf{x} = \{x_0, x_1, ..., x_{\hat{k}-1} \}$ has been generated, it must be interleaved. 
The interleaved (permuted) bits are then elaborated by the accumulator PPM (APPM) block, also named \textit{inner code}, composed by an accumulator and a memoryless PPM modulator. The accumulator can be described as a rate-1 code with transfer function $1/(1+D)$. 
Finally, \textbf{b} passes through the PPM modulator. At this point, the codeword is fractioned into $S=\hat{k}/m$ symbols, where $m=log_2(M)$ and $M$ the PPM order. The output of the PPM block is a slotted symbol sequence $\textbf{c}=\{ c_0^{q_0}, c_1^{q_1}, ..., c_{S-1}^{q_{S-1}} \}$, where each $\textbf{c}_i^{q_i}$ represents a PPM symbol with a laser pulse in the $q_i$ position. Vector $\textbf{q}=\{ q_0, q_1, ..., q_{S-1}\}$ defines the integer position values in the range between $0$ and $M-1$. 

The decoding is implemented as an iterative procedure very similar to the BCJR decoding \cite{scppm}. Considering the modulation and coding technique as a single large encoder and describing the respective trellis diagrams, it is possible to use a turbo-iterative demodulator and decoder.
This approach, proposed in \cite{scppm}, originates from turbo codes applied to the PPM channel and from iterative decoding, showing near-capacity performance of this decoding procedure.

\begin{figure}[h]
\begin{center}
\includegraphics[width=0.9\textwidth]{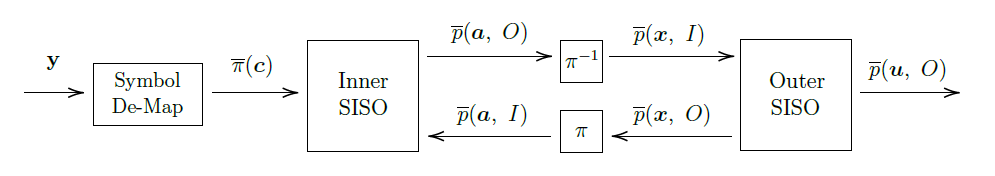}
\caption{Conceptual scheme of the SCPPM decoder.}\label{fig:decoder}
\end{center}
\end{figure}

Fig. \ref{fig:decoder} shows the block diagram of the SCPPM decoder architecture. 
As it can be seen, the demodulator system consists of two soft-input soft-output (SISO) blocks that mutually exchange soft information. Conceptually, they are the same and they are named \textit{inner} and \textit{outer} to differentiate the trellis description for the inner and outer code, respectively. An interleaver and a de-interleaver connect the flow of information between the two SISO blocks, and a symbol de-mapper converts the received photons in soft information. The soft information exchanged is the log-likelihood ratio (LLR) of two probabilities for the generic binary symbol $s_i$ defined as
\begin{equation}
\bar{\pi}(s_i) = log \frac{p_0(s_i)}{p_1(s_i)},
\label{eq:llr}
\end{equation}
where $p_0(s_i)$ and $p_1(s_i)$ represent the probabilities that $s_i$ is $0$ or $1$, respectively.

In the iterative decoding of serially concatenated codes, the extrinsic information from the inner SISO is fed to the outer SISO as a priori information. For the improvement in the error rate, iteration after iteration, the output mutual information from the outer SISO must be greater than the input mutual information to the inner SISO.
In \cite{Barsoum}, it has been discussed that the use of an accumulator just before the PPM modulation (i.e., APPM) improves the performance, with respect to a pure PPM.

\begin{figure}[h!]
\begin{center}
\includegraphics[width=0.75\textwidth]{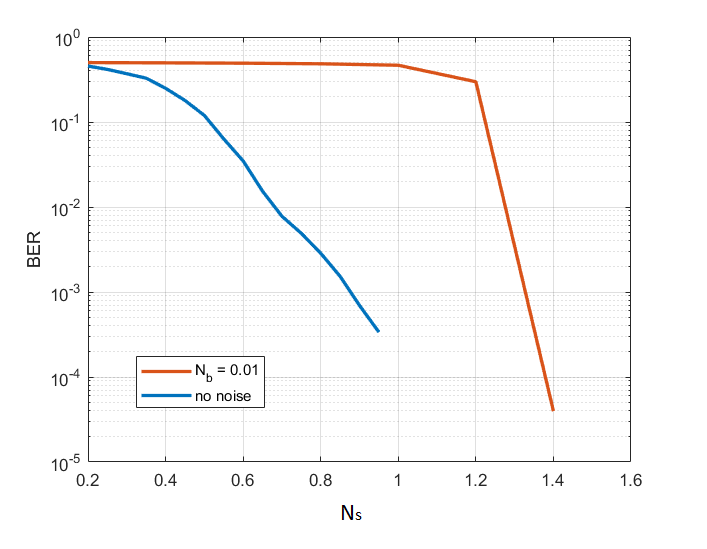}
\caption{SCPPM performance with code rate $r=1/3$, M=1024, $N_s$ in the range [0.2 1.6], and $N_b \in \{0, 0.01 \}$.}\label{fig:scppmBER}
\end{center}
\end{figure}

In terms of BER, Fig.~\ref{fig:scppmBER} shows the error rate as a function of $N_s$ for two values of $N_b$.

In the previous Section it is shown how the number and the positions of the emitters (intended as the leaves of the tree of light) affect the amount of received photon rate. Tab.~\ref{Tab:resQC} shows some results with an average transmitted power of $1$\,W. 

\begin{table}[h!] \renewcommand\arraystretch{1.4}
\centering
\begin{small}
\begin{tabular}{cccc}
\hline
{\# of sources} & {$P_{rx}$ [W]}  & {$n_s$ [ph/s]} & {\makecell{time to collect $N_{s, min}$\\ for a coded BER$\approx 10^{-2}$ [s]}}\\
\hline
1 &  $5.8\times10^{-25}$  & $2\times 10^{-6}$ & $7\times 10^{5}$ \\
3 &  $1.7\times10^{-24}$  &$7\times 10^{-6}$ & $2\times 10^{5}$  \\
9 &$5.2\times10^{-24}$  & $2\times 10^{-5}$ & $7\times 10^{4}$  \\
25 & $1.4\times10^{-34}$  & $6\times 10^{-5}$ & $2.33\times 10^{4}$ \\
64 & $3.7\times10^{-23}$   & $1\times 10^{-4}$ & $1.4\times 10^{4}$  \\
100 & $5.8\times10^{-23}$  & $2\times10^{-4}$ & $7\times 10^{3}$\\
625 & $3.6\times10^{-22}$   &$1\times 10^{-3}$ & $1.4\times 10^{3}$ \\
2500 &  $1.4\times10^{-21}$ &$6\times 10^{-3}$ & $0.23\times 10^{3}$ \\ 
10000 & $5.8\times10^{-21}$ & $0.02$ & $70$  \\
\hline
\end{tabular}
\end{small}
\caption{Link budget considerations.}
\label{Tab:resQC}
\end{table}

Merging the results of Fig.~\ref{fig:scppmBER} and Tab.~\ref{Tab:resQC} some considerations can be drawn. Assume that our objective is a $BER \approx 10^{-2}$ with SCPPM with code-rate 1/3 and background noise $N_b=0.01$. The minimum $N_s$ required is $\approx 1.4$. If we consider 10000 sources, the rate of received photons is 0.02\,Hz, and, consequently, we need to wait $70$\,s in order to collect $1.4$ useful photons. In this scenario, if the payload of useful bits to be transmitted is approximately $5$\,kbit, using 1024-PPM and therefore $10$\,bits per PPM symbol, with a guard factor $\alpha_{gt}=2.2s$ as in \cite{Messerschmitt_2020}, the number of required PPM symbols is $1512$. Then the time needed to collect the required number of useful photons to guarantee the objective BER is $231000$\,s. It is worth noting that this highly depends on the background noise, since $N_b$ strongly affects BER as shown in Fig.~\ref{fig:scppmBER}. 
This gets worst if we decrease the number of leaves of the ToL: the lower the number of leaves, the longer the time to collect the required photons (see  Tab.~\ref{Tab:resQC}). 

\begin{figure}[h]
\centering\includegraphics[width=0.72\textwidth]{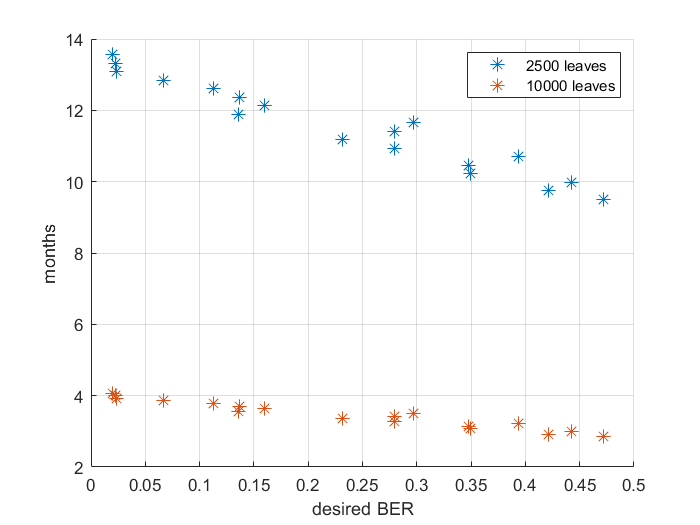}
\caption{Time to download an image of 1\,Mbit, using SCPPM with code rate 1/3, $M=1024$, $N_b = 0.01$. The number of leaves of the ToL is 2500 or 10000.}
\label{fig:bervsmonths}
\end{figure}

Fig.~\ref{fig:bervsmonths} shows the number of months needed to download an image of 1\,Mbit as a function of the desired BER, for 2500 or 10000 leaves of the ToL. It has been considered $N_b = 0.01 $. 


For high background noise values, other channel coding strategies should be considered.
One idea could be to decrease even further the code rate $r$ in order to improve the BER performance of Fig.~\ref{fig:scppmBER}. However, this decreases the net bit-rate (i.e., the useful bit rate exclusive of error correction), with a larger downloading time. Furthermore, the computation complexity and the time for the SCPPM decoder increase as well. 

To deal with this, LDPC codes and a symbol message passing (SMP) decoder for non-binary LDPC codes could be a promising solution. 

\subsection{Channel coding techniques for the Poisson channel - LDPC }\label{codingLDPC}

\begin{figure}[h!]
\begin{center}
\includegraphics[width=0.88\textwidth]{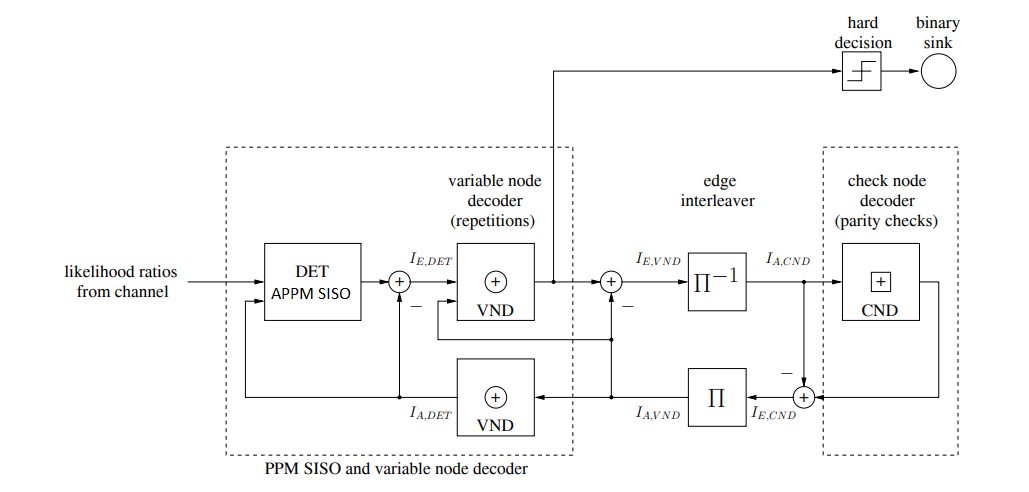}
\caption{Receiver model showing message passing flow and points for monitoring mutual information.}\label{fig:ldpcDec}
\end{center}
\end{figure}

LDPC codes are binary linear block codes where the codeword $\textbf{c}$ of $n$ bits is obtained from the input sequence $\textbf{u}$ of $k$ bits as $\textbf{c} = \textbf{u}\, \textbf{G}$, where the $k\times n$ matrix $\mathbf{G}$ is the code generator matrix. The block diagram of the LDPC decoder is shown in Fig.~\ref{fig:ldpcDec}, illustrating the flow of messages (likelihoods) passed between an LDPC decoder and an APPM SISO and within the LDPC decoder itself. 

LDPC codes can be defined by an $C$x$V$ sparse parity-check matrix $\mathbf{H}=[h_{i,j}]$ which can be represented by a Tanner graph with $V$ variable nodes (VNs) corresponding to the codeword symbols and $C$ check nodes (CNs) corresponding to parity checks. Each edge connecting the VN $v$ to the CN $c$ is labeled by a non-zero element $h_{v,c}$ of $\mathbf{H}$. The sets $N(v)$ and $N(c)$ denote the neighbors of VN $v$ and CN $c$, respectively. The degree of a VN $d_v$ is given by the cardinality of $N(v)$; similarly, the degree of a CN $d_c$ is the cardinality of $N(c)$. The code rate is given by $r=1-d_v/d_c$.
Similarly to the analysis of SCPPM, in which two decoders (inner and outer) exchange soft information, also the LDPC decoding process can be seen as the message-passing between variable nodes and check nodes, where the input of each step is taken as  \textit{a-priori information} and it is mapped into an output \textit{extrinsic information}.
In the literature \cite{Brink1, Brink2}, it has been proposed to track the evolution of the mutual information between bits and their corresponding LLRs to predict the decoder behavior. This can be done by means of the \textit{extrinsic information transfer} (\textit{EXIT}) \textit{function}.


The \textit{EXIT} function is a plot of the mutual information $I(A;L_e)$, where $A$ represents the bits and $L_e$ the extrinsic LLR,  as a function of $I(A;L)$, where $L$ are the a-priori LLR. 

\begin{figure}[h!]
     \centering
     \begin{subfigure}[b]{0.45\textwidth}
         \centering
         \includegraphics[width=\textwidth]{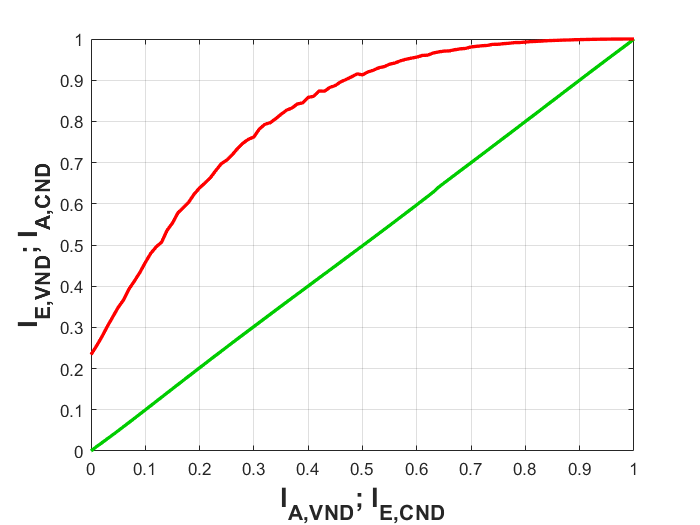}
         \caption{$N_s=0.5$.}
         \label{fig:fig2a}
     \end{subfigure}
     \hfill
     \begin{subfigure}[b]{0.45\textwidth}
         \centering
         \includegraphics[width=\textwidth]{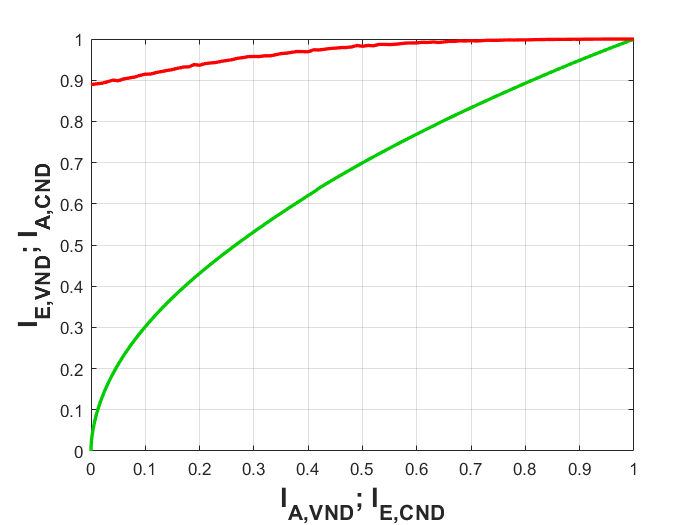}
         \caption{$N_s=3.5$.}
         \label{fig:fig2b}
     \end{subfigure}
     \hfill
        \caption{VND and CND EXIT curves with $M=1024$, $N_b=0.01$,  $d_v=2$, $d_c=3$ (i.e. code rate $\frac{1}{3}$).}
        \label{fig:fig2}
\end{figure}

\begin{figure}[h!]
     \centering
     \begin{subfigure}[b]{0.45\textwidth}
         \centering
         \includegraphics[width=\textwidth]{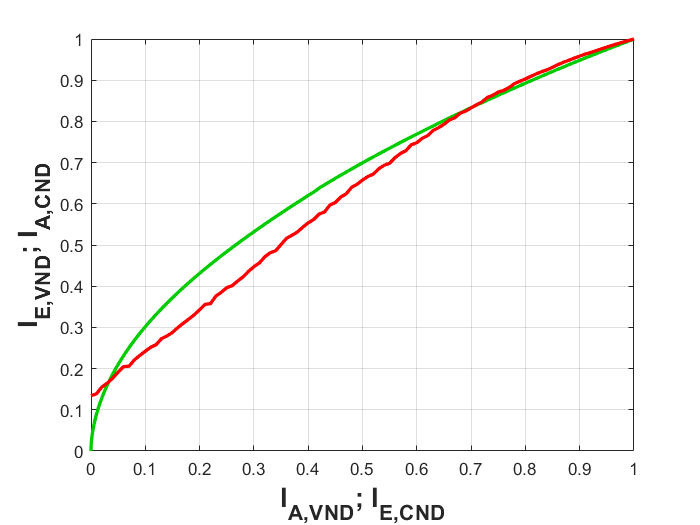}
         \caption{$N_s=0.5$.}
         \label{fig:fig3a}
     \end{subfigure}
     \hfill
     \begin{subfigure}[b]{0.45\textwidth}
         \centering
         \includegraphics[width=\textwidth]{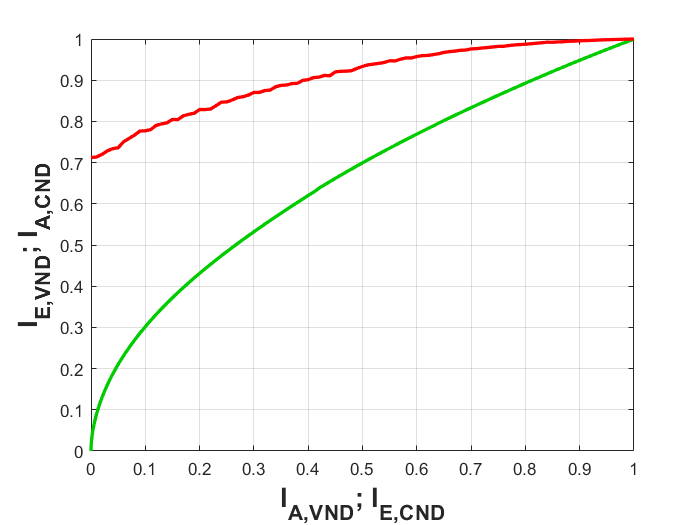}
         \caption{$N_s=3.5$.}
         \label{fig:fig3b}
     \end{subfigure}
     \hfill
        \caption{VND and CND EXIT curves with with $M=1024$, $N_b=0.1$, $d_v=2$, $d_c=3$ (i.e., code rate $\frac{1}{3}$).}
        \label{fig:fig3}
\end{figure}

\begin{figure}[h!]
     \centering
     \begin{subfigure}[b]{0.45\textwidth}
         \centering
         \includegraphics[width=\textwidth]{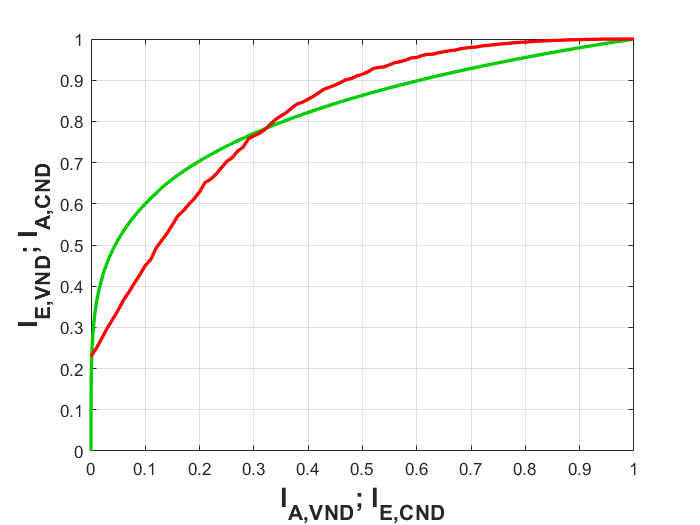}
         \caption{$N_s=0.5$.}
         \label{fig:fig5a}
     \end{subfigure}
     \hfill
     \begin{subfigure}[b]{0.45\textwidth}
         \centering
         \includegraphics[width=\textwidth]{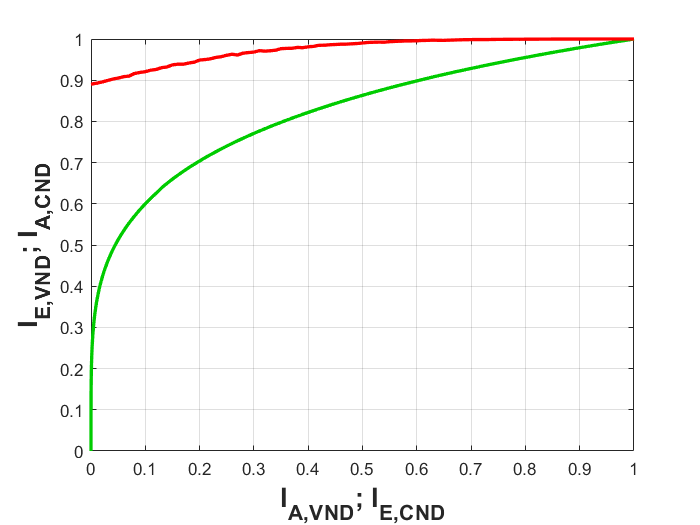}
         \caption{$N_s=3.5$.}
         \label{fig:fig5b}
     \end{subfigure}
     \hfill
        \caption{VND and CND EXIT curves with $M=1024$, $N_b=0.01$, $d_v=3$, $d_c=6$ (i.e. code rate $\frac{1}{2}$).}
         \label{fig:fig5}
\end{figure}

\begin{figure}[h!]
     \begin{subfigure}[b]{0.45\textwidth}
         \centering
         \includegraphics[width=\textwidth]{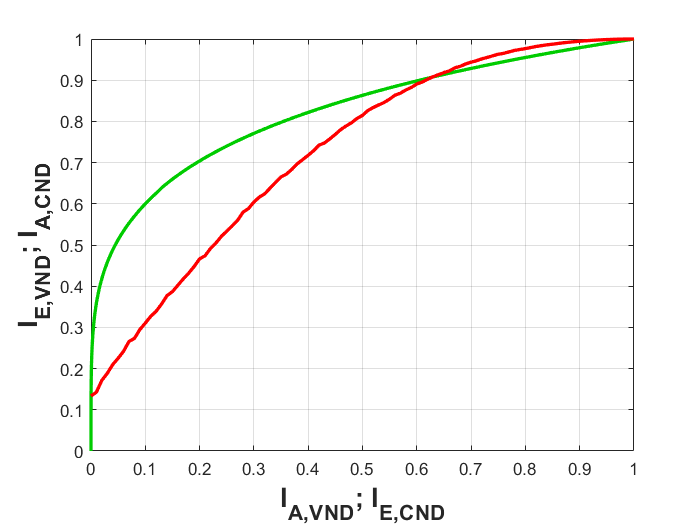}
         \caption{$N_s=0.5$.}
         \label{fig:fig6a}
     \end{subfigure}
     \hfill
     \begin{subfigure}[b]{0.45\textwidth}
         \centering
         \includegraphics[width=\textwidth]{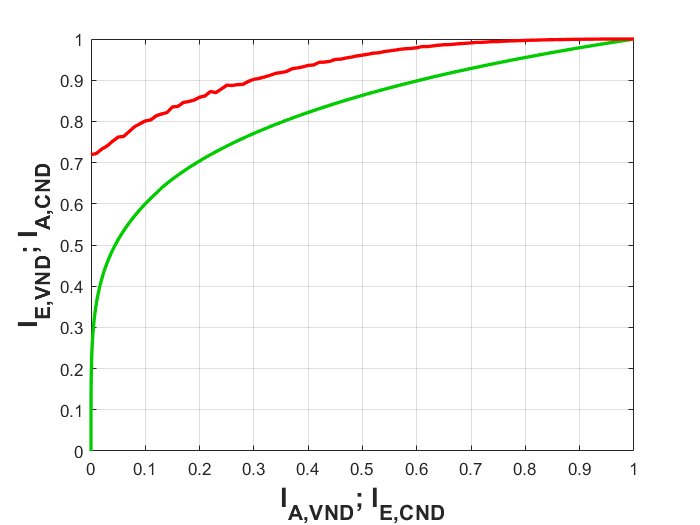}
         \caption{$N_s=3.5$.}
         \label{fig:fig6b}
     \end{subfigure}
     \hfill
        \caption{VND and CND EXIT curves curves with $M=1024$, $N_b=0.1$, $d_v=3$, $d_c=6$ (i.e., code rate $\frac{1}{2}$).}
        \label{fig:fig6}
\end{figure}

The EXIT analysis might be used to design the proper LDPC \cite{brink3}. 
We can determine an EXIT curve for the variable nodes of the LDPC decoder, denoted by \textit{VND}, and another one for the check nodes, \textit{CND}. 
For a regular LDPC code, \textit{the VND curve should lie above its corresponding  CND curve}.
Figs. \ref{fig:fig2}--\ref{fig:fig6} show the VND (red) curves and CND (green) curves. In particular, the red curves represent $I_{E, VND}$ as a function of $I_{A, VND}$, whereas the green curves show $I_{A, CND}$ as a function of $I_{E, CND}$. 
Consequently, in the cases in which the red curve is above the green curve the decoder should work properly. 
In the following results, we consider a PPM order $M=1024$ in all cases, signal photons $N_s=\{0.5, 3.5\}$ and noise photons $N_b=\{0.01, 0.1\}$.
Figs.~\ref{fig:fig2}--\ref{fig:fig3} refer to $d_v=2$, $d_c=3$ (i.e. code rate  $r=1/3$).

On the other hand if a higher code rate $1/2$ is considered, by setting $d_v=3$, $d_c=6$, the performance is even worse, as shown by the EXIT curves of Figs.~\ref{fig:fig5}--\ref{fig:fig6}. 

It is clear from the figures that the higher the background noise, the nearer the red and green curves are, up to a point that they cross and iterative decoding does not decrease the error rate. This is emphasized for low values of $N_s$ (e.g., $N_s=0.5$ $N_b=0.1$ of Figs.~\ref{fig:fig3a} and \ref{fig:fig6a}).

\clearpage
\subsection{Symbol message passing (SMP) non-binary LDPC for the Poisson channel}
\label{sec:smpldpc}

Non-binary LDPC codes are defined over a field $\mathbb{F}_q = \{ 0, 1, \alpha, .., \alpha^{q-2} \}$ with $q=2^m$, $m$ a positive integer and $\alpha$ a primitive element. The finite field order $q$ is matched to the PPM order yielding a one-to-one mapping between the symbol $c_i$ of the codeword and PPM symbols $\textbf{x} = (x_{i,1}, x_{i,2}, ..., x_{i,q})$.


Similarly to binary LDPC, non-binary LDPC codes can be defined by an $M$x$N$ sparse parity-check matrix $\mathbf{H}=[h_{i,j}]$ where in this case the matrix elements $h_{i,j}$ belong to $\mathbb{F}_q$. $\mathbf{H}$ is associated to a Tanner graph with $N$ variable nodes (VNs) representing the codeword symbols and $M$ check nodes (CNs) corresponding to parity checks.
The set $N(v)$ and $N(c)$ denote the neighbors of VN $v$ and CN $c$, respectively. The degree of a VN is given by the cardinality of $N(v)$. In a similar way, the degree of of a CN $c$ is the cardinality of $N(c)$. The VN (CN) edge-oriented degree distribution polynomial is $\lambda(x)= \sum_{i}{\lambda_i x^{i-1}}$ $\left( \rho(x)= \sum_{i}{\rho_i x^{i-1}} \right)$, where $\lambda_i$ ($\rho_i$) is the fraction of edges incident to VNs (CNs) with degree $i$. An unstructured irregular LDPC code ensemble $\mathcal{C}_{\lambda, \rho}^{q, N}$ is the set of all $q$ary LDPC codes with block length N and degree distribution polynomial pair $\lambda(x)$ and $\rho(x)$.

In the SMP decoding algorithm each VN $v$ computes the log-likelihood vector and sends the symbol which has the maximum L-value to all its neighbors, while the message from CN $c$ to its neighbor VN $v$ is obtained by determining the symbol that satisfies the parity-check equation. Symbol Message Passing algorithm can be extended by including erasures, symbol and erasure message passing (SEMP). 

Some proposals have been presented in \cite{smpLDPC2022}, where the authors designed optimized rate 1/2 irregular LDPC code ensembles for $q \in \{4,8,16,32\}$, $N_b \in \{0.002, 0.1\}$ for both SMP and SEMP decoding, showing that SMP/SEMP decoding might be a good choice when low-complexity decoding is targeted. 

However, in our scenario $q$ should assume much higher values than those considered in \cite{smpLDPC2022}, i.e., $q=1024, 2048$, etc. In these cases, an efficient way to evaluate the density evolution should be designed and only some upper and lower bounds can be obtained \cite{smpLDPC2019}.

\subsection{On the effective channel coding for the sails}
\label{subsec:futwork}
By considering the results described above, the channel coding for error protection on the Poisson PPM channel can then be based on  the following conclusions:
\begin{enumerate}
    \item The SCPPM coding scheme achieves a better performance  with respect to  Reed-Solomon codes, especially when the background noise is not negligible. 
    \item The EXIT function analysis can be used for the design of channel coding techniques, such as SCPPM and LDPC. 
    \item Symbol message passing (SMP) decoders for non-binary LDPC codes over the $q$-ary channel (with $q$ matched to the Poisson PPM channel) can be a promising channel coding technique.
\end{enumerate}
    %

\clearpage

\section{Conclusions}

With this study, we have   elaborated on the feasibility of the Starshot sail communication system from the physical, optical and communication theory point of views. 

Our approach on the optical realization considers the system as a Tree-of-Light that is realized on the surface of the sail by means of beam splitters, modulators, waveguides and grating couplers. In this way the transmitter functionality would allows keeping the mass of the sail low and may benefit on the scalability for the replication of the surface optics already demonstrated in current applications.

To tackle the extreme conditions of the channel, with the unprecedented link length and the corresponding optical losses, the study is proposing a beam emission that is based on the interference by an array of grating couplers that compresses the lobe to one microradian. In this way, a realistic laser source and the enhancement of the peak intensity due to the PPM protocol may allow for a sensible counting rate at the ground receiver. 

We have pointed out the criteria to be used to design the array for the beam combining and for the beam steering with active OPA technology. The divergence of the source is controlled by the array layout, specifically, the width of the beam in the far field scales inversely with the width of the array. We also proved that the intensity peak depends on the effective area of the transmitter. Therefore, if properly phased in coherence, $N$ sources give rise to an equivalent intensity peak to a single laser source with the same area, but they produce a much smaller beam spot at the receiver due to the power redistribution in side lobes. By the way, the shrinking of the main lobe is not a restrictive aspect as long as, after propagation, the beam results much larger than any feasible receiver which is then approximately illuminated by a constant intensity, the on-axis value in case of perfect pointing.

In addition, only an active phased array can provide beam steering without mechanical actuators, thus allowing to reduce the size, weight and power consumption of the sailcraft transmission system. The complexity of the electro-optical control system rises clearly with the number of emitters so that implementations with larger size GCs, arranged fewer in number, are convenient. 

From these considerations, it follows that most of the efforts has to be focused on the pointing capability, which comprises both a precise evaluation of the sailcraft-Earth relative position and a consistent steering of the Far-field distribution of the transmitted light. The latter implies that the correct calibration of the phase modulators is also a fundamental and delicate task, as well as the evaluation and eventual correction of any deformations in the sail that could affect the OPA spatial distribution.

The synchronization of the PPM protocol has been proposed to be realized by means of the modulation of polarization of the light for particular PPM symbols. The implementation of this modulation may be realized with a separate array for each state or with a suitable active grating coupler technology.

The design of the individual grating coupler to enhance the efficiency of conversion of the guided mode of the light carried by the waveguides to the free-space mode has reached the high level of 0.7, also addressing the polarization transformation into circular polarized state, well suited for preventing the misalignment issue between transmitter and receiver.

As pointed out above, optical communications over space channels are commonly designed as pulse-position-modulated (PPM) laser links and  one of our goals was to show the relevance of channel coding techniques to get as closer as possible to the channel capacity. Indeed, without channel coding, the source bits are directly mapped into PPM symbols and at the receiver side the maximum likelihood decoding, using the photodetector counts as observables, requires a maximum count selection for each PPM frame. If the resulting error probability is not low enough, coding must be used to improve performance, with the source bits first encoded into channel words, then the words sent to the PPM modulator. Indeed, the role of error correcting codes is to reconstruct a highly reliable replica of the scientific data transmitted by the sails. 

However, when channel coding is considered, it is no longer obvious that the maximum likelihood frame decisioning approach is optimal, since it does not allow for channel symbol erasures. For low levels of background noise, it has been argued that matched Reed Solomon (RS) coding appears as a natural encoding scheme, especially if only erasures can occur (since RS decoding has maximal capability for correcting erasures). 
On the other hand, when background is present in a non-negligible way, conversion of counts to channel symbols would involve errors as well as erasures.
Indeed, it has been shown that RS performance on a noisy PPM channel typically remains far away from capacity when conventional hard-decision decoding is used. As a consequence, we discussed the SCPPM scheme which refers to a precise combination of modulation and coding technique mostly used in a deep-space optical link scenario because of its characteristics that suits well for this kind of environment. The name SCPPM derives from the combination of a PPM modulator with other blocks serially concatenated to it. Among them, a convolutional code has been considered as the error-control code. 
Promising results have been found even in noisy scenario. Specifically, we found a BPP of $2.381 bit/ph$ for a background noise level of $0.01$ by means of a code rate $1/3$ convolutional code. However, SCPPM performance could be still improved both as BER results and computational resources requirements, and this could be one of the main purposes of a possible phase two of the Starshot project.

In conclusion, we are proposing here the ideas and the methods  that may be exploited to develop an experimental prototype for the sail transmission, coding and nano optics parts. 

Finally, we would like to warmly thank the Breakthrough Initiatives and in particular Dr. S. Pete Worden and James Schalkwyk,  for giving us the opportunity and the support to study this subject. A particular thanks goes to Professor Philip Mauskopf, of Arizona State University and Starshot Communications Research Director, for the great help in understanding  the context and the  spirit of creativity that he had shared with us.


\clearpage

\begin{thebibliography} {99}

\bibitem{Udry7} S. Udry et al., “The HARPS search for southern extra-solar planets,” Astron. Astrophys., vol. 469, no. 3, pp. L43–L47, Jul. 2007.

\bibitem{BIStarshot}
https://breakthroughinitiatives.org/initiative/3

\bibitem{Forw} R. L. Forward, Roundtrip interstellar travel using laser-pushed lightsails Journal of Spacecraft and Rockets. American Institute of Aeronautics and Astronautics (AIAA) 21 187 (1984)

\bibitem{Messerschmitt_2020}
D. G. Messerschmitt, P. Lubin, and I. Morrison, 
``Challenges in Scientific Data Communication from Low-mass Interstellar Probes,'' 
{\it The Astrophysical Journal Supplement Series}, Vol.~249, No.~2, 2020. 

\bibitem{Villoresi2008} P. Villoresi et al., “Experimental verification of the feasibility of a quantum channel between space and Earth,” New J. Phys., 10 033038, 2008.

\bibitem{BCFig} A. Berera and J. Calderón-Figueroa, “Viability of quantum communication across interstellar distances,” Phys. Rev. D 105 123033 (2022)


\bibitem{Guo2021} Guo Y., Guo Y., Li C., Zhang H., Zhou X., Zhang L., "Integrated Optical Phased Arrays for Beam Forming and Steering", {\it Applied Sciences}, 2021, 11(9):4017, doi: 10.3390/app11094017

\bibitem{Wolfgang1994} Wolfgang M. Neubert, Klaus H. Kudielka, Walter R. Leeb, and Arpad L. Scholtz, ``Experimental demonstration of an optical phased array antenna for laser space communications,''{\it Appl. Opt.} 33, 3820-3830, 1994.


\bibitem{Vasey1993}F. Vasey, F. K. Reinhart, R. Houdré, and J. M. Stauffer, "Spatial optical beam steering with an AlGaAs integrated phased array," Appl. Opt. 32, 3220-3232 (1993)

\bibitem{VanAcoleyen2009} K. Van Acoleyen, W. Bogaerts, J. Jágerská, N. Le Thomas, R. Houdré, and R. Baets, "Off-chip beam steering with a one-dimensional optical phased array on silicon-on-insulator," {\it Opt. Lett.} 34, 1477-1479 (2009)

\bibitem{Guerber2022} S. Guerber, D. Fowler, J. Faugier-Tovar, K. Abdoul Carim, B. Delplanque, and B. Szelag, "Wafer-level calibration of large-scale integrated optical phased arrays," {\it Opt. Express} 30, 35246-35255 (2022)



\bibitem{Zhang21} M. Zhang, C. Wang, P. Kharel, D. Zhu, and M. Lončar, “Integrated lithium niobate electro-optic modulators: when performance meets scalability,” Optica, 8 652 (2021)



\bibitem{He19} M. He et al., High-performance hybrid silicon and lithium niobate Mach-Zehnder modulators for 100 Gbit/s and beyond, Nat. Photonics, 13 359 (2019)

\bibitem{Feng10} N.N. Feng et al., High speed carrier-depletion modulators with 14V-cm $V_{\pi}$L integrated on 0.25 $\mu$m silicon-on-insulator waveguides, Opt. Express, 18 7994 (2010)

\bibitem{Mar} R. Marchetti, C. Lacava, L. Carroll, K. Gradkowski, and P. Minzioni, Coupling strategies for silicon photonics integrated chips [Invited], Photonics Res. 7 201 (2019)

\bibitem{Vallone15} G. Vallone et al., “Experimental Satellite Quantum Communications,” {\sl Phys. Rev. Lett.}, vol. 115, no. 4, p. 040502, Jul. 2015.

\bibitem{Liao17} S.-K. Liao et al., “Satellite-to-ground quantum key distribution,” {\sl Nature}, vol. 549, no. 7670, pp. 43–47, Aug. 2017.

 

\bibitem{xu} Xu, J., Yang, S., Wu, L., Xu, L., Li, Y., Liao, R., ... \& Cheng, X. (2021). Design and fabrication of a high-performance binary blazed grating coupler for perfectly perpendicular coupling. {\it Optics Express}, 29(26), 42999-43010.
%
\bibitem{taillaert} Taillaert, D., Bienstman, P., and Baets, R. (2004). Compact efficient broadband grating coupler for silicon-on-insulator waveguides. {\it Optics letters}, 29(23), 2749-2751.
%
\bibitem{zadka} Zadka, M., Chang, Y. C., Mohanty, A., Phare, C. T., Roberts, S. P., and Lipson, M. (2017, May). Millimeter long grating coupler with uniform spatial output. In {\it 2017 Conference on Lasers and Electro-Optics (CLEO)} (pp. 1-2). IEEE.
\bibitem{kim} Kim, S., Westly, D. A., Roxworthy, B. J., Li, Q., Yulaev, A., Srinivasan, K., and Aksyuk, V. A. (2018). Photonic waveguide to free-space Gaussian beam extreme mode converter. {\it Light: Science \& Applications}, 7(1), 72.
\bibitem{marchetti} Marchetti, R., Lacava, C., Khokhar, A., Chen, X., Cristiani, I., Richardson, D. J., ... and Minzioni, P. (2017). High-efficiency grating-couplers: demonstration of a new design strategy. {\it Scientific reports}, 7(1), 16670.
\bibitem{Khorasaninejad} Khorasaninejad, M. \& Crozier, K. B. Silicon nanofin grating as a miniature chirality-distinguishing beam-splitter. {\it Nat. Commun.} 5(1), 1-6 (2014).
\bibitem{Zheng} Zheng, G., Mühlenbernd, H., Kenney, M., Li, G., Zentgraf, T. \& Zhang, S. Metasurface holograms reaching 80\% efficiency. {\it Nat. Nanotech.} 10(4), 308-312 (2015).
\bibitem{Vogliardi} Vogliardi, A., Romanato, F. \& Ruffato, G. Design of Dual-Functional Metaoptics for the Spin-Controlled Generation of Orbital Angular Momentum Beams. {\it Front. Phys.} 586 (2022).
\bibitem{Vogliardi2} Vogliardi, A., Ruffato, G., Dal Zilio, S., Bonaldo, D., \& Romanato, F. Dual-functional metalenses for the polarization-controlled generation of focalized vector beams in the telecom infrared. {\it Scientific Reports}, 13(1), 10327 (2023).
%
\bibitem{Zhang} Zhang, K., Yuan, Y., Ding, X., Ratni, B., Burokur, S. N. \& Wu, Q. High-efficiency metalenses with switchable functionalities in microwave region. {\it ACS Appl. Mater. Interfaces} 11(31), 28423-28430 (2019). 
%
\bibitem{McManamon1996} P. F. McManamon \textit{et al.} "Optical phased array technology," in \textit{Proceedings of the IEEE}, vol. 84, no. 2, pp. 268-298, Feb. 1996, doi: 10.1109/5.482231.
%
\bibitem{Heck2017} Heck, Martijn J.R.. "Highly integrated optical phased arrays: photonic integrated circuits for optical beam shaping and beam steering", \textit{Nanophotonics} 6, no. 1 2017: 93-107. doi: 10.1515/nanoph-2015-0152.
%



\bibitem{scppm}
B. Moision,  ``Coded modulation for the deep space optical channel: serially concatenated PPM",  
{\it IPN Progress Report}, vol. 42, no. 162, 2005.
%
\bibitem{Barsoum}
M. F. Barsoum et al.,  "``EXIT function aided design of iteratively decodable codes for the Poisson PPM channel",  
{\it IEEE Trans. Commun.}, vol. 58, no. 12, pp. 3573--3582, 2010.
%
\bibitem{brink3}
S. Ten Brink et al.,  ``Design of low-density parity-check codes for modulation and detection",  
{\it IEEE Trans. Commun.}, vol. 52, no. 4, pp. 670--678, 2004.
%
\bibitem{Brink1}
S. Ten Brink,  "``Convergence of iterative decoding",  
{\it Electron. Lett.}, vol. 35, no. 10, pp. 806--808, 1999.
%
\bibitem{Brink2}
S. Ten Brink,  ``Convergence behaviour of iteratively parallel concatenation codes",  {\it IEEE Trans. Commun.}, vol. 49, no. 10, pp. 1727--1737, 2001.
%
%
\bibitem{smpLDPC2022}
E. B. Yacoub et al., ``Analysis of symbol message passing LDPC decoder for the Poisson PPM channel", IEEE International Symposium on Information Theory (ISIT), 2022.
%
\bibitem{smpLDPC2019}
F. Lazaro et al., ``Symbol message passing decoding of nonbinary low-density parity-check codes", IEEE Globecom, 2019.
%

\end{thebibliography}
\end{document}